\documentclass[fleqn,usenatbib]{mnras}
\usepackage[T1]{fontenc}
\usepackage{ae,aecompl}
\usepackage{graphicx}
\usepackage{amsmath}
\usepackage{amssymb}
\usepackage{txfonts}

\title[Expected IMBHs: Late-type galaxies]{Expected intermediate mass black
  holes in the Virgo cluster. II. Late-type galaxies}

\author[Graham et al.]{
Alister W.\ Graham$^{1}$\thanks{E-mail: AGraham@swin.edu.au},
Roberto Soria$^{2,3,4}$, 
and Benjamin L.\ Davis$^{1}$
\\
$^{1}$Centre for Astrophysics and Supercomputing, Swinburne University of
Technology, Hawthorn, VIC 3122, Australia.\\
$^{2}$College of Astronomy and Space Sciences, University of the Chinese
Academy of Sciences, Beijing 100049, China\\  
$^{3}$International Centre for Radio Astronomy Research, Curtin University, GPO Box U1987, Perth, WA 6845, Australia.\\
$^4$Sydney Institute for Astronomy, School of Physics A28, The University of
Sydney, Sydney, NSW 2006, Australia.
}

\date{Accepted XXX. Received YYY; in original form ZZZ}
\pubyear{2018}

\begin{document}
\label{firstpage}
\pagerange{\pageref{firstpage}--\pageref{lastpage}}
\maketitle

\begin{abstract}

The {\it Chandra X-ray Observatory's} Cycle~18 Large Program titled `Spiral
galaxies of the Virgo Cluster' will image 52 galaxies with the ACIS-S
detector. Combined with archival data for an additional 22 galaxies, this 
will represent the complete sample of 74 spiral galaxies in the Virgo cluster
with star-formation rates $\gtrsim$0.3\,$M_{\odot}$~yr$^{-1}$.  
Many of these galaxies are expected to have an active nucleus,
signalling the presence of a central black hole.  In preparation for this survey,
we predict the central black hole masses using the 
latest black hole scaling relations based on spiral arm pitch angle $\phi$,
velocity dispersion $\sigma$, and total stellar mass $M_{\rm *,galaxy}$.  
With a focus on intermediate mass black holes ($10^2<M_{\rm
  bh}/M_{\odot}<10^5$), we highlight NGC~4713 and NGC~4178, both with $M_{\rm
  bh}\approx10^3$--$10^4$ (an estimate which is further supported in NGC~4178
by its nuclear star cluster mass). 
From {\it Chandra} archival data, we find that both galaxies have a point-like nuclear X-ray source, 
with unabsorbed 0.3--10 keV luminosities of a few times 10$^{38}$~erg~s$^{-1}$. 
In NGC~4178, the nuclear source has a soft, probably thermal, spectrum 
consistent with a stellar-mass black hole in the high/soft state, while no strong constraints 
can be derived for the nuclear emission of NGC~4713. 
In total, 33 of the 74 galaxies are predicted to have $M_{\rm
  bh}<(10^5$--$10^6)\,M_{\odot}$, and several are consistently 
predicted, via three methods, to have masses of
$10^4$--$10^5\,M_{\odot}$, such as IC~3392, NGC~4294 and NGC~4413.  We
speculate that a sizeable 
population of IMBHs may reside in late-type spiral galaxies with low stellar
mass ($M_{\ast}\lesssim10^{10}\,M_{\odot}$).

\end{abstract}

\begin{keywords}
black hole physics -- 
X-rays: galaxies --
(galaxies:) quasars: supermassive black holes -- 
galaxies: spiral --
galaxies: structure -- 
galaxies: individual: IC~3392, NGC~4178, NGC~4294, NGC~4413, NGC~4470, NGC~4713
\end{keywords}

\section{Introduction}\label{SecIntro}

In the young universe, massive metal-free Population III stars (Schwarzschild
\& Spitzer 1953; Larson 1998) may have spawned `intermediate mass black holes' 
(IMBHs) with masses greater than $10^2\, M_{\odot}$ (e.g.\ Bond et al.\ 1984;
Carr et al.\ 1984; Madau \& Rees 2001; Schneider et al.\ 2002), but see Umeda
\& Nomoto (2003) and Fraser et al.\ (2017) who cap the `Pop III' masses at
120-130 $M_{\odot}$.  Additional mechanisms have also been proposed for the
creation of IMBHs (see, e.g., Miller \& Colbert 2004 and Mezcua 2017),
including: the runaway merging of stellar mass black holes and stars
(Zel'dovich \& Podurets 1965; Larson 1970; Shapiro \& Teukolsky
1985; Quinlan \& Shapiro 1990; Portegies Zwart \& McMillan 2002; G\"urkan et
al.\ 2004); primordial black holes (e.g.\ Argyres et al.\ 1998; Bean \&
Magueijo 2002; Carr et al.\ 2010; Grobov et al.\ 2011); the direct collapse of
massive gas clouds, bypassing the Pop~III stage 
(Doroshkevich et al.\ 1967; Umemure et al.\ 1993; Bromm \& 
Loeb 2003; Mayer et al.\ 2010); and a stunted or inefficient growth of nuclear
black holes via gas accretion at the centres of galaxies (e.g.\ Johnson \&
Bromm 2007; Sijacki et al.\ 2007; Alvarez et al.\ 2009; Heckman \& Best
2014). In the last of those alternative scenarios, IMBHs are an intermediate
step on the way to the maturation of supermassive black holes (SMBHs,
$M_{\rm bh} > 10^5\, M_{\odot}$; Rees 1984; Shankar et al.\ 2004; Ferrarese \&
Ford 2005; Kormendy \& Ho 2013; Graham 2016a, and references therein).

In contrast to the plethora of theoretical formation models, direct
observational detection of IMBHs remains elusive. There is a long history of
disproved suggestions and claims of IMBHs in globular clusters, stretching
back to at least the X-ray data from Clark et al.\ (1975).  Most recently, the
presence of an IMBH with a mass of $\approx$2000\,$M_{\odot}$ in the core of
the Milky Way globular cluster 47 Tuc was suggested by a kinematic modelling
of its pulsars (Kiziltan et al.\ 2017), but there is no electromagnetic evidence
for its existence, nor proof of any other IMBH in Galactic globular clusters
(Anderson \& van der Marel 2010; Strader et al.\ 2012). 

In the centre of nearby galaxies, there are only a handful of candidate IMBHs
with an X-ray detection, i.e.\ with plausible signature of gas accretion onto
a compact object. These include: NGC\,4178\footnote{For NGC\,4178, the
  prediction that $M_{\rm bh} < 10^5\, M_{\odot}$ is simply based on the
  assumption that the nuclear BH mass is less than 20 per cent of this
  galaxy's nuclear star cluster mass.} (Satyapal et al.\ 2009; Secrest et
al.\ 2012); LEDA\,87300 (Baldassare et al.\ 2015; Graham et al.\ 2016);
NGC\,404 (Nguyen et al.\ 2017); NGC~3319 (Jiang et al.\ 2018); and possibly
NGC\,4395 (Iwasawa et al.\ 2000; Shih et al.\ 2003; Filippenko \& Ho 2003,
Nucita et al.\ 2017, but see den Brok et al.\ 2015).

Outside of galactic nuclei, IMBH searches initially focused on a rare class of
point-like X-ray sources with X-ray luminosities $\sim$10$^{40}$--$10^{41}$
erg s$^{-1}$ (e.g.\ Colbert \& Mushotzky 1999; Swartz et al.\ 2008; Sutton et
al.\ 2012; Mezcua et al.\ 2015; Zolotukhin et al.\ 2016). This was partly
based on the assumption that the X-ray luminosity of an accreting compact
object cannot be much in excess of its classical Eddington limit (hence,
luminosities $\ga$10$^{40}$ erg s$^{-1}$ would require BH masses $\ga$100\,
$M_{\odot}$), and partly on the detection of a low-temperature thermal
component ($kT \sim 0.2$ keV) that was interpreted as emission from an IMBH
accretion disk (Miller et al.\ 2003).  However, most of the sources in this
class are today interpreted as super-Eddington stellar-mass black holes or
neutron stars (Feng \& Soria 2011; Kaaret et al.\ 2017).  To date, the most
solid IMBH identification in this class of off-nuclear sources is HLX-1, in
the galaxy cluster Abell 2877, and seen in projection near the S0 galaxy
ESO\,243-49 (Farrell et al.\ 2009, Soria et al.\ 2010; Yan et al.\ 2015; Webb
et al.\ 2010, 2017). HLX-1 has a mass of $\sim$10$^4$\,$M_{\odot}$ (Davis et
al.\ 2011; Godet et al.\ 2012; Soria et al.\ 2017) and may reside in the
remnant nucleus of a gravitationally-captured and tidally-stripped satellite
galaxy (Mapelli et al.\ 2013; Farrell et al.\ 2014), which leads us back to
galactic nuclei as the most likely cradle of IMBHs.

In this work, we focus on IMBH candidates in galactic nuclei. Due to their
low mass, it is currently impossible to spatially resolve the gravitational
sphere-of-influence of these black holes; therefore, astronomers need to rely
on alternative means to gauge their mass.  There are now numerous galaxy
parameters that can be used to predict the mass of a galaxy's central black
hole, and Koliopanos et al.\ (2017) report on the consistency of various black
hole scaling relations.  

The existence, or scarcity, of central IMBHs obviously has implications for
theories regarding the growth of supermassive black holes. For example, some
have theorised that supermassive black holes started from seed masses
$\ga$10$^5\, M_{\odot}$ --- created from the direct collapse of large gas
clouds and viscous high-density accretion-discs (e.g.\ Haehnelt \& Rees 1993;
Loeb \& Rasio 1994; Koushiappas, Bullock \& Dekel 2004; Regan et al.\ 2017)
--- which could potentially bypass the very existence of IMBHs.  Therefore,
defining the demography of IMBHs has implications for the co-evolution of
massive black holes and their host galaxy alike.

For two reasons, spiral galaxies may represent a more promising field to
plough than early-type galaxies or dwarf galaxies\footnote{ Given the rarity
  of dwarf spiral galaxies (Schombert et al.\ 1995; Graham et al.\ 2003),
  dwarf galaxies are overwhelmingly early-type galaxies.}.  This is due to
their low mass bulges and disks --- and thus low mass black holes --- and the
presence of gas which may result in an active galactic nucleus around the
central black hole, potentially betraying the black hole's presence.  Until
very recently, the largest sample of spiral galaxies, with directly measured
BH masses, that had been carefully decomposed into their various structural
components, e.g.\ bar, bulge, rings, etc., and therefore with reliable bulge
parameters, stood at 17 galaxies (Savorgnan \& Graham 2016).  This has now
more than doubled, with a sample of 43 such spiral galaxies\footnote{With a
  central rather than global spiral pattern, we exclude the ES galaxy Cygnus~A
  from the list of 44 galaxies in Davis et al.\ (2017), who, we note, reported
  that three of these remaining 43 galaxies appear to be bulgeless.}
presented in Davis et al.\ (2018a), along with revised and notably more
accurate $M_{\rm bh}$--$M_{\rm *,bulge}$ and $M_{\rm bh}$--$M_{\rm *,galaxy}$
relations for the spiral galaxies (Davis et al.\ 2018b).  

Here, we apply three independent, updated, black hole scaling relations to a
sample of 74 spiral galaxies in the Virgo cluster.  X-ray images already exist
for 22 members of this sample, and new images will be acquired for the
remaining members during the {\it Chandra X-ray Observatory's} Cycle~18
observing program (see Section~\ref{SecData}).  This paper's tabulation of
predicted black hole masses for these 74 galaxies will serve as a reference, enabling
two key objectives to be met. First, in the pursuit of evidence for the (largely)
missing population of IMBHs, we will eventually be able to say which of the 74
galaxies predicted to have an IMBH additionally contain electromagnetic
evidence for the existence of a black hole. We are not, however, just laying
the necessary groundwork for this, but we are able to now, and do, explore
which of the initial 22 galaxies contain both an active galactic 
nucleus (AGN) and a predicted IMBH.
Second, by combining the existing and upcoming X-ray data with the predicted
black hole masses for the full sample, we will be able to compute the black
holes' Eddington ratios and investigate how the average Eddington-scaled X-ray
luminosity scales with BH mass (Soria et al.\ 2018, in preparation). Gallo et
al.\ (2010) have already attempted this measurement for the early-type
galaxies in the Virgo cluster, and in 
Graham \& Soria (2018, hereafter Paper~I) we 
revisit this measurement using updated black hole
scaling relations for early-type galaxies, such that in low-mass systems the black hole mass
scales quadratically, rather than linearly, with the early-type galaxies'
$B$-band luminosity (Graham \& Scott 2013). 

The layout of this current paper is as follows.  In Section~(\ref{SecData}) 
we briefly introduce the galaxy set that will be analysed.  A more complete
description will be provided in Soria et al.\ (2018, in preparation). In
Section~(\ref{sec_Param}) we explain the measurements of pitch angle,
velocity dispersion, and stellar mass that we have acquired for these 74
galaxies, and we introduce the latest (spiral galaxy) black hole scaling
relations involving these quantities, from which
we derive the expected black hole masses, that are  
presented in the Appendix.  In
Section~(\ref{Sec_Results}) we compare the black hole mass predictions from the
three independent methods.  We additionally take the opportunity to combine
the black hole scaling relations by eliminating the black hole mass term and
providing revised galaxy scaling relations between pitch angle, velocity
dispersion, and galaxy stellar mass.  In Section~(\ref{Sec_X}) we pay
particular attention to galaxies predicted to have black hole masses less than
$10^5\, M_{\odot}$, and we investigate the X-ray properties of those nuclei for
which archival X-ray data already exists.  Finally, Section~(\ref{Sec_Disc})
provides a discussion of various related issues.

\section{Galaxy Sample}\label{SecData}

Soria et al.\ (2018, in preparation) selected the complete sample of 74 Virgo
cluster spiral galaxies with star-formation rates $>$0.3\,$M_{\odot}$
yr$^{-1}$ (see the Appendix for this galaxy list).  This resulted in a mix of
(early- and late-type) spiral galaxies, in the inner and outer regions of the
cluster, spanning more than 5 mag in absolute $B$-band magnitude from roughly
$-$18 to $-$23 mag (Vega).  
Of these 74 galaxies, just three have directly measured black hole masses; they are:
NGC~4303, $\log(M_{\rm bh}/M_{\odot}) = 6.58^{+0.07}_{-0.26}$ (Pastorini et al.\ 2007); 
NGC~4388, $\log(M_{\rm bh}/M_{\odot}) = 6.90^{+0.04}_{-0.05}$ (Tadhunter et al.\ 2003); 
and NGC~4501, $\log(M_{\rm bh}/M_{\odot}) = 7.13^{+0.08}_{-0.08}$ (Saglia et al.\ 2016).

In the X-ray bands, 22 of those galaxies already have archival {\it Chandra
  X-ray Observatory} data, and the rest are currently being observed with the
Advanced CCD Imaging Spectrometer (ACIS-S) detector, as part of a 559-ks {\it
  Chandra} Large Project titled `Spiral galaxies of the Virgo cluster' (PI:
R.~Soria.  Proposal ID: 18620568).  General results for our X-ray study,
(including both nuclear and non-nuclear source catalogues, luminosity
functions, multiband identifications, and comparisons between the X-ray
properties as a function of Hubble type, will be presented in forthcoming
work, once the observations have been completed.  Here, we only use the
archival {\it Chandra} data to characterise the nuclear X-ray properties of
spiral galaxies that we identify as possible IMBH hosts, based on their black
hole scaling relations.

\section{Predicting black hole masses}\label{sec_Param}

In this section, we introduce the three\footnote{There is also a scaling
  relation between $M_{\rm bh}$ and the bulge S\'ersic index $n$ (Graham \&
  Driver 2007; Savorgnan 2016; Davis et al.\ 2018a).  However, we do not use
  that relation for this work, partly because of the steepness at low masses,
  and partly to avoid the need for bulge/disk decompositions of our Virgo
  sample.}  black hole scaling relations that will be used to predict the
black hole masses of our Virgo cluster spiral galaxy sample, and we describe
where the three associated parameter sets came from.

\subsection{Pitch Angles}

For galaxies whose disks are suitably inclined, such that their spiral pattern
is visible, we project these images to a face-on orientation and measure their
spiral arm `pitch
angle' $\phi$, i.e.\ how tightly or loosely wound their spiral arms are.  
The mathematical description of the pitch
angle, and the method of image analysis, is detailed in Davis et al.\ (2017), which 
also presents a significantly updated $M_{\rm bh}$--$|\phi|$ relation 
(equation~(\ref{eq1a}), below) for spiral galaxies, 
building on Seigar et al.\ (2008) and Berrier et al.\ (2013).  

As noted in Davis et al.\ (2017), a prominent difficulty in pitch angle
measurement is the identification of the fundamental pitch angle, which is
analogous to the fundamental frequency in the musical harmonic series of
frequencies. Pitched musical instruments produce musical notes with a
characteristic timbre that is defined by the summation of a fundamental
frequency and naturally occurring harmonics (integer multiples of the
fundamental frequency). Careful Fourier analysis of the sound will allow
discovery of the fundamental frequency and any perceptible harmonics. A
synonymous scenario occurs in the measurement of galactic spiral arm pitch
angle via two-dimensional Fourier analysis (Kalnajs 1975; Iye et al.\ 1982;
Krakow et al.\ 1982; Puerari \& Dottori 1992; Seigar et al.\ 2005; Davis et
al.\ 2012; Yu et al.\ 2018). Therefore, pitch angle measurement methods, when
performed in haste, can incorrectly select a `harmonic' pitch angle instead of
the `fundamental' pitch angle.

Similarly, the Fourier analysis of sound becomes less certain when the source
tone is soft, short duration, or blended with contaminating noise. Spiral
galaxies also become more difficult to analyse when resolution is poor, their
disk orientation is close to edge-on, their spiral structure is intrinsically
flocculent, or the arc length of their spiral segments are short. Whereas the
former problems are stochastic and lead to increased uncertainty in pitch
angle measurements (i.e., constant mean with an increased standard deviation),
the latter problem of short spiral arc segments (i.e., small subtended polar
angle) poses a potential systematic bias and can lead one to incorrectly
identify a harmonic rather than the fundamental pitch angle. Typically, this
problem manifests itself when spiral arc segments subtend polar angles
$<\pi/2$ radians.

One clear benefit is that 
the measurement of galactic spiral arm pitch angle only requires simple
imaging that highlights a perceptible spiral pattern, without the need of any
photometric calibrations. Therefore, we accessed publicly available imaging
from telescopes such as the Galaxy Evolution Explorer (GALEX), 
Hubble Space Telescope (HST), 
Spitzer Space Telescope (SST), 
Sloan Digital Sky Survey (SDSS), 
etc. This wide selection of
telescopes also implies a wide range of passbands from far-ultraviolet up to
mid-infrared wavelengths. Pour-Imani et al.\ (2016) concluded that pitch angle
is statistically tighter in passbands that reveal young stellar populations,
such as ultraviolet filters. The difference between young stellar spiral
patterns and old stellar spiral patterns is small, typically less than 4
degrees in pitch angle. Because of this, we preferentially use young stellar
passbands when they are available and if the resolution is sufficient to
clearly display the spiral pattern. The same preference was applied in the
derivation of the $M_{\rm bh}$--|$\phi$| relation in Davis et al.\ (2017).

The bisector linear 
regression between black hole mass and the absolute value of the spiral arm
pitch angle,
for the full sample of 44 `spiral' galaxies\footnote{Davis et 
al.\ (2017) reported that excluding Cygnus~A from the linear regression
between black hole mass and spiral arm pitch angle did not have a significant
effect.} 
with directly measured black hole masses, is such that 
\begin{equation}\label{eq1a}
\log (M_{\rm bh}/M_{\odot}) = (7.01\pm0.07) - (0.171\pm0.017) ( |\phi^{\circ}| -
15^{\circ} ), 
\end{equation} 
with an intrinsic and total rms scatter in the $\log M_{\rm bh}$ direction of
$0.30\pm0.08$ and 0.43 dex, respectively (Davis et al.\ 2017, their equation~8).

Importantly, and curiously, the rms scatter in the $\log M_{\rm bh}$ direction
about this black hole scaling relation is smaller than the rms scatter
observed in the other black hole scaling relations.  This is in part due to
the shallow slope of the relation in equation~(\ref{eq1a}), and because of the
careful pitch angle measurements that were determined using three different
approaches (see Davis et al.\ 2017 for details).  In passing, we note that the
bulgeless galaxy NGC~2748 probably had an incorrect pitch angle assigned to
it.  Removing this galaxy, along with the early-type ES galaxy Cygnus~A, plus
two potential outliers (NGC 5055 and NGC 4395) seen in Davis et al.\ (2017,
their figure~4), gives the revised and more robust relation 
\begin{equation}\label{eq1b}
\log (M_{\rm bh}/M_{\odot}) = (7.03\pm0.07) - (0.164\pm0.018) ( |\phi^{\circ}|
-15^{\circ} ), 
\end{equation}
with intrinsic and total rms scatter equal to $0.31\pm0.07$ and 0.41 dex,
respectively. 
Equation~(\ref{eq1b}) has
been used here to predict the black hole masses in 43 Virgo cluster spiral
galaxies for which we were able to determine their pitch angle.  The results
are presented in the Appendix table.

\subsection{Velocity Dispersions}

Homogenised velocity dispersions are available in
Hyperleda\footnote{http://leda.univ-lyon1.fr} (Paturel et al.\ 2003) for 39 of the 74 Virgo
galaxies. 
We have assigned a 15\% uncertainty to each of these values.

The bisector linear regression between $\log M_{\rm
 bh}$ and $\log \sigma$ --- taken from Table~4 in Davis et al.\ (2017) for
their reduced sample of 40 spiral galaxies (see below) --- is given by 
\begin{eqnarray}\label{eq2}
\log (M_{\rm bh}/M_{\odot}) = (8.06\pm0.13) + \nonumber \\
   (5.65\pm0.79)\log \left( \sigma / 200\, {\rm km\, s}^{-1} \right). 
\end{eqnarray}
The intrinsic scatter is $0.51\pm0.04$ dex in the $\log M_{\rm bh}$ direction,
and the total rms scatter is 0.63 dex in the $\log M_{\rm bh}$ direction.  
The slope of this expression agrees well with the $M_{\rm bh}$--$\sigma$ relation from 
Savorgnan \& Graham (2015, their Table~2), who found that their bisector regression yielded
slopes between 4.8 and 5.7 for both `fast rotators' and `slow rotators'. 
We have used equation~(\ref{eq2}) to predict the black hole masses for those
39 Virgo galaxies with available velocity dispersions, and we provide these 
values in the Appendix table. 

As noted above, in deriving equation~(\ref{eq2}), four galaxies were excluded
from the initial sample of 44 galaxies with directly measured black hole
masses.  NGC~6926 has no reported velocity dispersion, while Cygnus~A is not a
(typical) spiral galaxy, but rather an ES galaxy (see the discussion in Graham et
al.\ 2016) with a nuclear, rather than 
large-scale, bar and spiral pattern.  Another such example, albeit in a dwarf
ES galaxy, is LEDA~2108986 (Graham et al.\ 2017).  Finally, NGC~4395 and NGC~5055 are
outliers that appear to have unusually low velocity dispersions; they were
also excluded by Davis et al.\ (2017) in order to obtain a more robust
regression unbiased by outliers.

\subsection{Galaxy Stellar Masses}

As revealed by the $M_{\rm bh}$--$|\phi|$ relation in Davis et al.\ (2017, see
also Seigar et al.\ 2008 and Ringermacher \& Mead 2009), the central black
hole masses in spiral galaxies are not unrelated to their disks.  Furthermore,
disks contain the bulk of the stellar mass in spiral galaxies.
 
\begin{figure}
	\includegraphics[trim=1cm 2cm 2cm 1.5cm, width=\columnwidth]{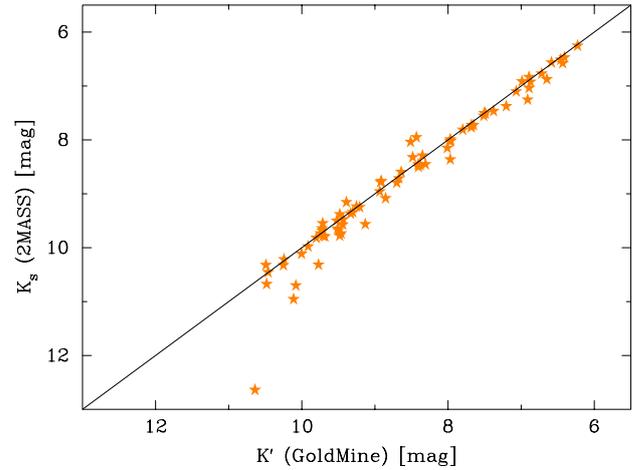}
    \caption{2MASS $K$-band apparent magnitudes versus the
      $K$-band magnitudes from GOLD Mine.  Both magnitudes have been
      corrected for Galactic extinction.  At the faint end, some of the 2MASS
      magnitudes under-estimate the galaxy light.} 
    \label{Fig0c}
\end{figure}

We have derived total galaxy stellar masses for our sample of 74 Virgo 
cluster galaxies via the $K^{\prime}$-band (2.2 $\mu$m) total apparent
magnitudes (Vega) available in the 
GOLD Mine\footnote{http://goldmine.mib.infn.it/} database (Gavazzi et
al.\ 2003). 
We had initially explored using the {\it Two Micron All Sky Survey}
(2MASS\footnote{www.ipac.caltech.edu/2mass}, Jarrett et al.\ 2000) $K_s$-band
total apparent magnitudes (Vega), but 
it sometimes under-estimates the galaxy luminosities (e.g., Kirby et
al.\ 2008; Schombert 2011), as can be seen in Figure~\ref{Fig0c}. 
The GOLD Mine apparent magnitudes were 
converted into absolute magnitudes using the mean, redshift-independent,
distance moduli provided by the NASA/IPAC Extragalactic Database
(NED)\footnote{http://nedwww.ipac.caltech.edu}. 
These absolute magnitudes were 
converted into solar units using an absolute magnitude for the Sun of
$\mathfrak{M}_{\odot ,K} = 3.28$ mag (Vega), taken from Willmer (2018), 
and then converted into a stellar mass, or rather a scaled-luminosity, using a
constant $K$-band stellar mass-to-light ratio $M/L_K=0.62$. 
The uncertainty that we have associated with our (GOLD Mine)-based stellar
masses --- 
which are tabulated in the Appendix --- stems from adding in quadrature: 
(i) an assumed 10 per cent error on the apparent stellar luminosity, 
(ii) the standard deviation provided by NED for the mean redshift-independent distance modulus;
and 
(iii) a 15 per cent error on the stellar mass-to-light ratio. 

\begin{figure}
	\includegraphics[trim=1cm 2cm 2cm 1.5cm,width=\columnwidth]{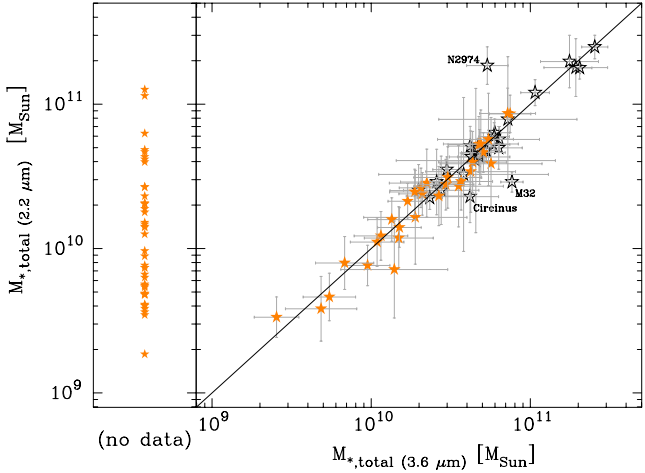}
    \caption{Galaxy stellar masses based on 2.2
      $\mu$m magnitudes (with no dust correction), and using $M/L_K = 0.62$, 
      versus the stellar masses based on the {\it Spitzer} 3.6
      $\mu$m magnitudes (using a constant (stellar mass)-to-(total light) ratio $M_*/L_{\rm tot}=0.45$, from
      Querejeta et al.\ (2015). For the Virgo sample (orange stars), for which we used the
      GOLD Mine $K$-band data, 31
      galaxies have Spitzer data.  For the 43 galaxies (excluding the Milky
      Way)  with directly measured black
      hole masses, we used the 2MASS $K$-band data (open black stars). 
      The $K$-band data for NGC~2974 is likely contaminated by a foreground star.
} 
    \label{Fig0a}
\end{figure}

We have been able to verify these stellar masses by using, when available, 
the published 3.6-$\mu$m {\it Spitzer} galaxy magnitudes.  Using the same 
redshift-independent distance moduli provided by NED\footnote{We note that
  NGC~4276 only has one redshift-independent distance estimate; and we
  hereafter use the (Virgo + Great-Attractor + Shapley)-infall adjusted
  distance from NED for this galaxy.}, Laine et al.\ (2014,
their table~1) provide absolute 
galaxy magnitudes (AB, not Vega), at 3.6 $\mu$m, for 31 of our 74 galaxies.  
On average, 25\% of a spiral galaxy's flux at 3.6 $\mu$m comes from the glow
of dust (Querejeta et al.\ 2015, their Figures~8 and 9).  We therefore dim
Laine et al.'s magnitudes by 25\% before converting them into stellar masses
using $M_{\odot,3.6}=6.02$ (AB mag) and a (stellar mass)-to-(stellar light)
ratio $M/L_{3.6}=0.60$ (Meidt et al.\ 2014)\footnote{Based on a Chabrier
  (2003) initial stellar mass function.}.  This (stellar mass)-to-(stellar
light) ratio, coupled with the above mentioned 25\% flux reduction due to
glowing dust, yields a (stellar mass)-to-(total light) ratio of 0.45, or
$\log(M_*/L_{\rm tot}) = -0.35$, which can be seen in Figure~10 of Querejeta
et al.\ (2015) to provide a good approximation for more than 1600 large and
bright nearby spiral galaxies.  A comparison of these 31 Spitzer-based 
stellar masses with our 
(GOLD Mine)-based stellar masses can be seen in Figure~(\ref{Fig0a}).  These
masses are better thought of as scaled-luminosities, and we will return to
this issue in the following subsection.

Using a symmetrical implementation\footnote{This involves taking the
  bisector of the `forward' and `inverse' regressions.} of the
 modified {\sc FITEXY} routine (Press et al.\ 1992) from Tremaine et
al.\ (2002), Davis et al.\ (2018b) reported a linear regression between 
black hole mass and galaxy stellar mass, 
for their sample of 40 spiral galaxies with directly measured black hole
masses (excluding the ES transition galaxy 
Cygnus~A, and the three bulgeless galaxies NGC~4395, NGC~2748 and NGC~6926). 
The relation is
\begin{eqnarray}\label{eq3} 
\log (M_{\rm bh}/M_{\odot}) =  (7.26\pm0.14) + \nonumber \\
     (2.65\pm0.65)\log 
\left[ M_{\rm *,galaxy} / \upsilon (6.37\times10^{10}\, M_{\odot}) \right], 
\end{eqnarray}
where $\upsilon$ (lowercase $\Upsilon$) 
is a corrective stellar mass-to-light ratio term ---  
which depends on the initial mass function of the stars and the star formation
history ---
(see Davis et al.\ 2018a) that we can set equal to 1 
given the agreement seen in Figure~\ref{Fig0a}.  The intrinsic scatter and
total rms scatter in the $\log M_{\rm bh}$ direction is equal to $0.64$ and
0.75 dex, respectively.  Equation~(\ref{eq3}) was used to predict the black
hole masses for our 74 spiral galaxies, and the results are tabulated in the
Appendix.

If the stellar mass is wrong by 50\%, then the predicted logarithm of the
black hole mass will be off by 0.47 dex.  Combining this offset with the
1$\sigma$ intrinsic scatter in equation~(\ref{eq3}), one could find that the
predicted black hole mass is $\sim$1 dex, i.e.\ an order of magnitude,
different from the actual black hole mass.  We therefore place less confidence
in the black hole masses predicted from only the galaxy stellar mass.
However, readers should be aware that our reported intrinsic and total rms
scatters are not error-weighted quantities. That is, they can be dominated by
outlying data points with large error bars, and therefore they can give a
misleading view of how tightly defined the scaling relations are.  The slope
and intercept of the scaling relations presented here, and their associated
uncertainty, do however take into account the error bars on the data used to
define them. 

Finally, Davis et al.\ (2018b) additionally reported the following steeper $M_{\rm
  bh}$--$M_{\rm *,galaxy}$ relation, derived using a sophisticated Bayesian
analysis, 
\begin{eqnarray}\label{eq3B}
\log (M_{\rm bh}/M_{\odot}) =  (7.25^{+0.13}_{-0.14}) + \nonumber \\
     (3.05^{+0.57}_{-0.49})\log 
\left[ M_{\rm *,galaxy} / \upsilon (6.37\times10^{10}\, M_{\odot}) \right]. 
\end{eqnarray} 
This relation predicts black masses which agree well with those predicted from
our $M_{\rm bh}$--$\sigma$ relation (equation~\ref{eq2}), but it tends to
yield lower black hole masses than those predicted from our $M_{\rm
  bh}$--$|\phi|$ relation (equation~\ref{eq1a}).  
This is also true for equation~\ref{eq3}, and will be quantifed in
Section~\ref{Sec_Results}. 
Erring on the side of 
caution, such that we do not want to under-estimate the black hole masses and
claim a greater population of IMBHs than actually exists, we proceed by using
Equation~(\ref{eq3}) as our primary black hole mass based on the galaxy
stellar mass.  Black hole masses based on equation~\ref{eq3B}  are, however,
additionally included.

\subsubsection{What about colour-dependent $M/L$ ratios?} 

We have assumed that the previous $M_{\rm bh}$--$M_{\rm *,tot}$ relations are log-linear,
and we extrapolate this to masses below that which was used to define them. 
However, given that some of our Virgo cluster spiral
galaxies are less massive and bluer than those in Davis et al.\ (2018b), it may be
helpful if we provide some insight into what happens if the scaling which 
gives the scaled-luminosity, i.e.\ the so-called stellar mass, is not constant. 

The 40 spiral galaxies used to define the above $M_{\rm bh}$--$M_{\rm *,tot}$
relations have stellar masses greater than $2\times 10^{10}\,M_{\odot}$ (and
absolute $K$-band magnitudes brighter than $\approx -23$ mag), and, therefore,
the assumption of a constant 3.6~$\mu$m $M_*/L_*$ ratio of 0.60 (and
$M_*/L_{\rm tot}$ ratio of 0.45) --- which was used to derive the stellar
masses in Davis et al.\ (2018b) --- is likely to be a good approximation.
This is because these galaxies' stellar populations have roughly the same red
colour.  As such, the $M_{\rm bh}$--$M_{\rm *,tot}$ relations from Davis et
al.\ (2018b) can be thought of as a (black hole)-(scaled luminosity) relation.
Had the Davis et al.\ (2018b) sample contained some less massive {\it blue}
galaxies, then, for the following reason, one may expect the $M_{\rm
  bh}$--luminosity relation not to be log-linear, but to steepen at the faint
end.

\begin{figure}
  \includegraphics[trim=2.0cm 2cm 8.6cm 1.5cm, width=\columnwidth]{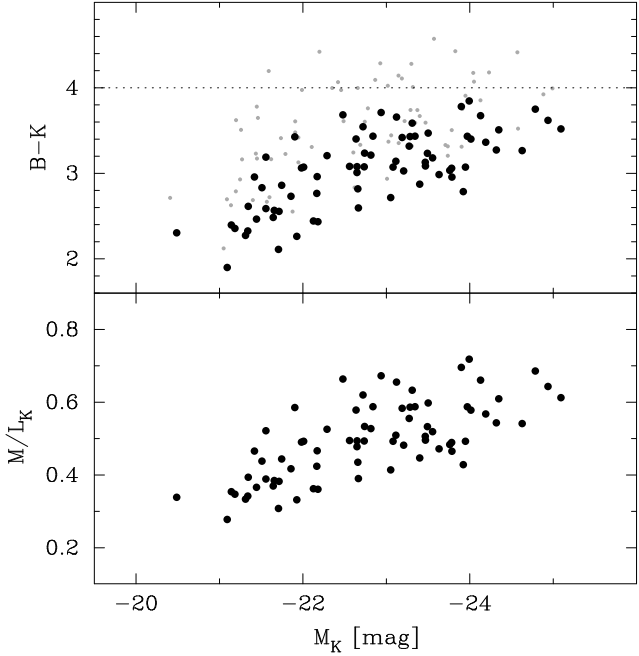}
    \caption{(RC3 $B$-band) - (Gold Mine $K^{\prime}$-band) colour, and the
      ($B-K^{\prime}$)-dependent $K^{\prime}$-band stellar mass-to-light
      ratio (equation~\ref{Eq_Bell}), versus the $K^{\prime}$-band absolute magnitude.  The grey points
    in the upper panel are based on the observed magnitudes, while the black
    points have been corrected for dust/inclination dimming using the prescription
    in Driver et al.\ (2008). 
}
    \label{FigB-K}
\end{figure}

Bell \& de Jong (2001) provide the following equation for the $K$-band
(stellar mass)-to-(stellar light) ratio as a function of the $B-K$
optical-(near-infrared) colour: 
\begin{equation}\label{Eq_Bell}
\log M/L_{K} = 0.2119(B-K)-0.9586.
\end{equation} 
We have obtained the 2MASS $K$-band 
data, and the RC3
 $B$-band data\footnote{The (Vega) $B$-band magnitudes are the $B_T$ values from the
  {\it Third Reference Catalogue of Bright Galaxies} (de Vaucouleurs et
  al.\ 1991) as tabulated in NED, and
  were subsequently corrected for Galactic extinction using the values from 
  Schlafly \& Finkbeiner 2011, also tabulated in NED.} 
, for our Virgo spiral galaxies.  Their $B-K$ colour, and the
associated $M/L_{K}$ ratio, is displayed in Figure~\ref{FigB-K}. 
One can see that at $\mathfrak{M}_K \gtrsim -23$ mag, the $M/L_{K}$ ratios
become smaller.  To maintain the log-linear $M_{\rm bh}$--$M_{\rm *,tot}$
relation (equation~\ref{eq3}),
obviously the $M_{\rm bh}$--luminosity relation needs to steepen for 
$\mathfrak{M}_K \gtrsim -23$ mag. 
If we were to employ the falling $M/L_{K}$ ratios seen in Figure~\ref{FigB-K}
as one progresses to fainter galaxies, then 
we would also need to employ this steeper $M_{\rm bh}$--$M_{\rm *,tot}$
relation at these magnitudes.  The net effect would be to cancel out
and return one to the single log-linear $M_{\rm bh}$--$M_{\rm *,tot}$
relation (equation~\ref{eq3}) that we are using together with a constant
$M/L_K=0.62$ for the GOLD Mine $K$-band data. 

\begin{figure}
\includegraphics[trim=1cm 2cm 2cm 1.5cm,width=\columnwidth]{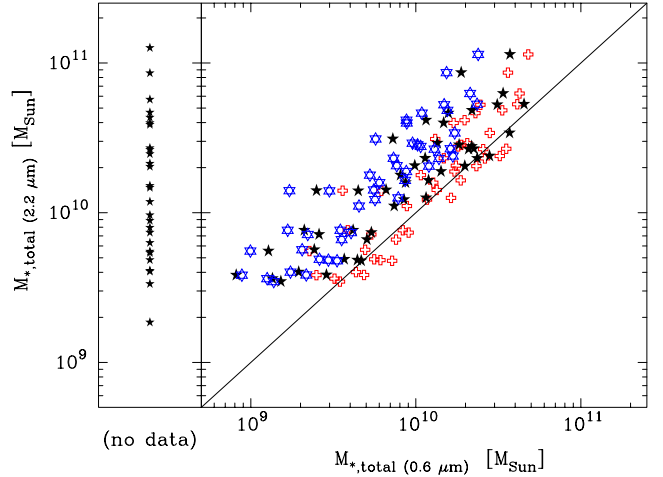}
\caption{Stellar masses based on the GOLD Mine 2.2-$\mu$m $K^{\prime}$-band magnitudes 
  (not dust corrected, and using $M/L_K=0.62$) 
  versus the
  stellar masses based on the observed (not dust corrected) 
  SDSS $i^{\prime}$-band 0.62-$\mu$m magnitudes 
  (using a [$g^{\prime}-i^{\prime}$]-dependent $M_*/L_{i^{\prime}}$ ratio)
  from Bell et al.\ (2003, red crosses), 
 Taylor et al.\ (2011, open blue hexagram), 
 and Roediger \& Courteau (2015, black filled stars).  
The data  reveal the need for the $\upsilon$ term in equations~\ref{eq3} and
  \ref{eq3B}.}
    \label{Fig0b}
\end{figure}

There is one additional element worthy of some exploration, and it pertains to
the $\upsilon$ term seen in equations~\ref{eq3} and \ref{eq3B}.
We have 
made use of the SDSS Data Release 12 (Alam et al.\ 2015) 
to obtain three additional stellar mass estimates.  Taylor 
et al.\ (2011) advocated that a ($g^{\prime}-i^{\prime}$)-dependent
$i^{\prime}$-band stellar mass-to-light ratio, $M_*/L_{i^{\prime}}$, yields
reliable stellar masses.  Their relation is such that 
$\log(M_*/L_{i^{\prime}})=0.70(g^{\prime}-i^{\prime})-0.68$, and applies to
the observed, i.e.\ not the dust-corrected, magnitudes.  We have also used the
relation $\log(M_*/L_{i^{\prime}})=0.518(g^{\prime}-i^{\prime})-0.152$ from
Bell et al.\ (2003).  Reddening due to dust will roughly move galaxies along
this relation (see Figure~6 in Bell et al.\ 2003, and Figure~13 in Driver et
al.\ 2007), and thus the relation can be applied to either the dust-corrected
or observed magnitudes; for consistency with Taylor et al.\ (2011), we have
chosen the latter.  Finally, based on the stellar population synthesis model
of Conroy et al.\ (2009), Roediger \& Courteau (2015) give the relation
$\log(M_*/L_{i^{\prime}})=0.979(g^{\prime}-i^{\prime})-0.831$.  These three
relations for the mass-to-light ratios have given us three more sets of
stellar mass estimates for (most of) our 74 spiral galaxies,
which are shown in Figure~\ref{Fig0b} against the (GOLD Mine
$K^{\prime}$)-based mass estimates.  While small 
random differences are apparent, due to uncertainties in the
magnitudes and 
simplicities in the stellar population models, the main offsets that are
visible can be captured / expressed by the $\upsilon$ term. 

\section{Results}
\label{Sec_Results}

\begin{figure}
	\includegraphics[trim=1cm 2cm 2cm 1.5cm, width=\columnwidth]{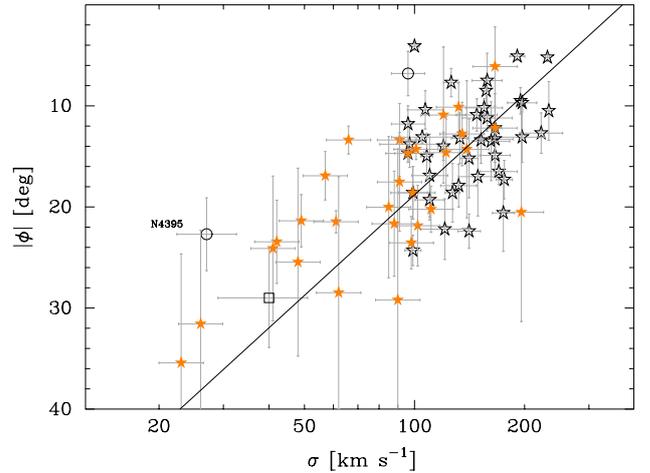}
    \caption{Absolute value of the spiral arm pitch angle, $|\phi|$, versus
      the stellar velocity dispersion $\sigma$.  The open black stars (and
      circles) represent spiral galaxies with bulges (and without bulges) that
      have directly measured black hole masses (see Davis et al.\ 2017).  The
      line represents equation~(\ref{eq4}), and is the expected trend based on
      the $M_{\rm bh}$--$|\phi|$ relation given in equation~(\ref{eq1a}) and
      the $M_{\rm bh}$--$\sigma$ relation given in equation~(\ref{eq2}) for
      galaxies with directly measured black hole masses.  The filled orange
      stars are the Virgo cluster spiral galaxies studied in this work.  They
      appear to follow the line well, as does LEDA~87300 (open square).} 
    \label{Fig3}
\end{figure}

\begin{figure}
	\includegraphics[trim=1cm 2cm 2cm 1.5cm, width=\columnwidth]{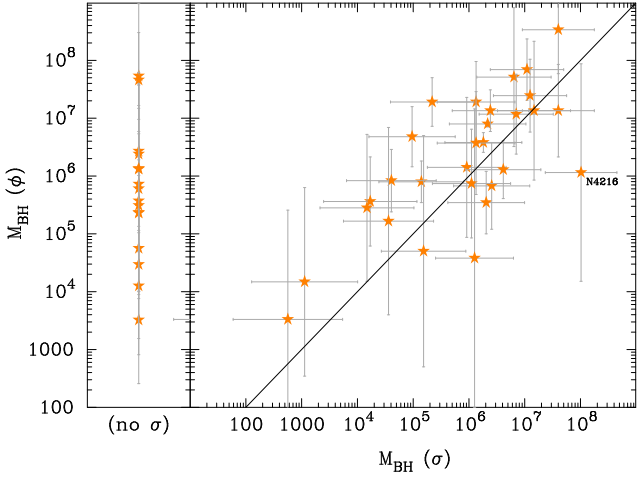}
    \caption{The predicted black hole masses in our Virgo galaxy sample,
      derived from, when available, the absolute value of the pitch angle
      $|\phi|$ (using equation~(\ref{eq1a})) and the velocity dispersion
      $\sigma$ (using equation~(\ref{eq2})).}
    \label{Fig4}
\end{figure}

\subsection{Pitch Angle vs Velocity Dispersion}

Combining equations~(\ref{eq1a}) and (\ref{eq2}) to eliminate $M_{\rm bh}$, one
obtains the relation
\begin{equation}
\log(\sigma/200\, {\rm km\, s}^{-1} ) = 0.268 - 0.030|\phi|. 
\label{eq4}
\end{equation}
This is shown by the line in Figure~(\ref{Fig3}), which plots $|\phi|$
versus $\log \sigma$ for spiral galaxies with directly measured black hole masses,
plus our sample of Virgo cluster spiral galaxies, NGC~4395 and
LEDA~87300. 

The Virgo galaxies appear consistent with the trend (equation~(\ref{eq4})) 
defined by the galaxy 
sample with directly measured black hole masses. Using equations~(\ref{eq1a})
and (\ref{eq2}) to predict the black hole masses in these Virgo galaxies, we
plot the results in Figure~(\ref{Fig4}). 
Of particular interest are NGC~4178 (Secrest et al.\ 2012) and NGC~4713, 
the two galaxies in the lower
left of the right hand panel, plus NGC~4294 in the lower 
section of the left hand panel.  They are predicted here to 
have black hole masses of $10^3$ to $10^4 \,M_{\odot}$ (see
the Appendix for every galaxies' predicted BH mass).

\begin{figure}
	\includegraphics[trim=1cm 2cm 2cm 1.5cm, width=\columnwidth]{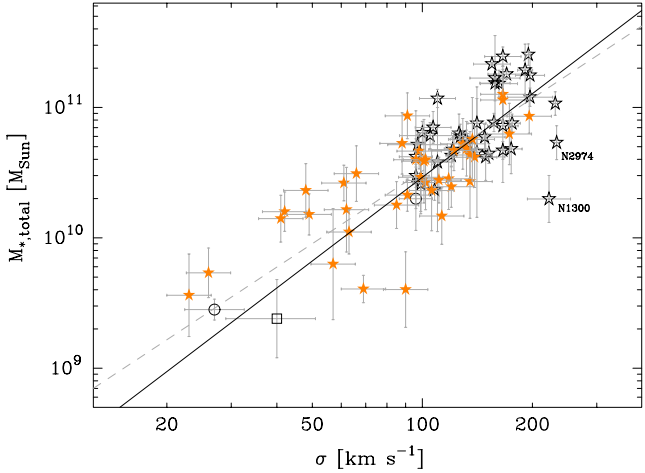}
    \caption{Galaxy stellar mass versus stellar velocity dispersion.  Note:
      neither LEDA~87300 (open square) nor NGC~4395 (lower left open circle)
      were used in either the linear regression between $M_{\rm bh}$ and
      $\sigma$, nor between $M_{\rm bh}$ and $M_{\rm *,galaxy}$, for the
      galaxy set with directly measured black hole masses (open stars and
      circles).  Those regressions (equations~(\ref{eq2}) and (\ref{eq3}),
      respectively), have been combined to produce equation~(\ref{eq5}) which
      is shown here by the solid line which has a slope of 2.13.  The bulk of the
      Virgo cluster spiral galaxies (filled orange stars) appear to follow
      this line well.  Equation~\ref{eq5B}, constructed from
      equation~\ref{eq2} and equation~\ref{eq3B}, is shown by the dashed grey
      line and has a slope equal to 1.85. 
    }
    \label{Fig2}
\end{figure}

\begin{figure}
	\includegraphics[trim=1cm 2cm 2cm 1.5cm, width=\columnwidth]{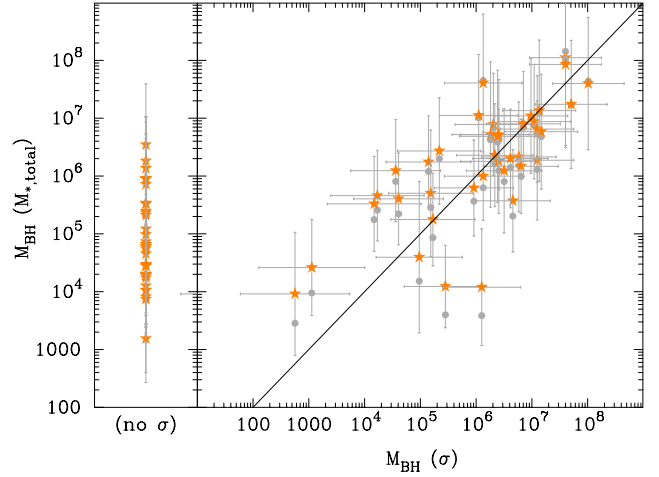}
    \caption{The orange stars show the predicted black hole masses in our Virgo galaxy sample,
      derived from, when available, the galaxy's stellar mass $M_{\rm
        *,galaxy}$ (using equation~(\ref{eq3})) and the stellar velocity
      dispersion $\sigma$ (using equation~(\ref{eq2})).  The grey circles had 
      their ($M_{\rm *,total}$)-based black hole masses derived using equation~\ref{eq3B}.
}
    \label{Fig1}
\end{figure}

\subsection{Stellar Mass vs Velocity Dispersion}

Combining equations~(\ref{eq2}) and (\ref{eq3}) to eliminate $M_{\rm bh}$, one
obtains the relation
\begin{equation}
\log ( M_{\rm *,galaxy} / 6.37\times10^{10}\, M_{\odot} ) = 0.302 + 2.132\log(\sigma/200\,
     {\rm km\, s}^{-1} ). 
\label{eq5}
\end{equation}
This is shown by the solid line in Figure~(\ref{Fig2}), which is a plot of
$M_{\rm *,galaxy}$ 
versus $\log \sigma$ for spiral galaxies with directly measured black hole masses, and
for our sample of Virgo cluster spiral galaxies. 

In Figure~(\ref{Fig1}), we display the result of using equations~(\ref{eq2})
and (\ref{eq3}) to predict the black hole masses in our Virgo galaxy sample. 
As before, two galaxies stand out, they are NGC~4178 and NGC~4713, the two galaxies in the
lower left of the right hand panel of Figure~(\ref{Fig1}). In addition, we note NGC~4396 and
NGC~4299 in the lower section of the left hand panel of Figure~(\ref{Fig1})

Coupling equation~\ref{eq3B}, rather than equation~\ref{eq3}, with
equation~\ref{eq2} results in the relation 
\begin{equation}
\log ( M_{\rm *,galaxy} / 6.37\times10^{10}\, M_{\odot} ) = 0.266 +
1.852\log(\sigma/200\, {\rm km\, s}^{-1} ).
\label{eq5B}
\end{equation}
This equation is represented by the 
dashed grey line in Figure~\ref{Fig2}. 
Equation~\ref{eq5} and \ref{eq5B} give the scaling relation $M_{\rm *,galaxy}
\propto \sigma^{2\pm0.15}$ for spiral galaxies, which matches well with the
relation for dwarf and ordinary early-type galaxies fainter than $\mathfrak{M}_B \approx
-20.5$ mag (e.g.\ Davies et al.\ 1983; Matkovi\'c \& Guzm\'an 2005).

\begin{figure}
	\includegraphics[trim=1cm 2cm 2cm 1.5cm, width=\columnwidth]{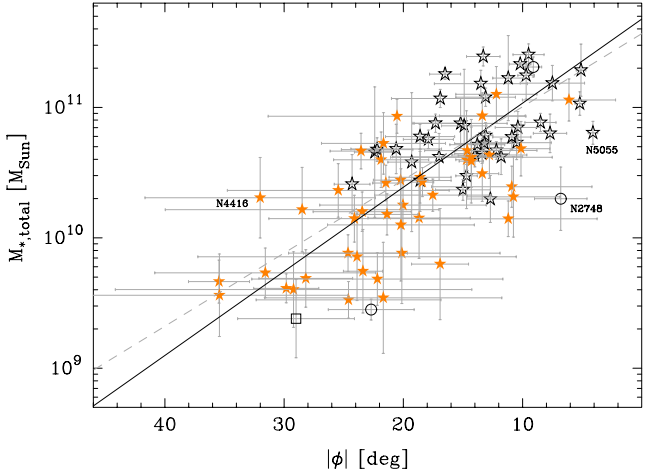}
    \caption{Galaxy stellar mass versus the absolute value of the spiral arm
      pitch angle.  Symbols have the same meaning as in Figure~(\ref{Fig3}).
      The solid line represents equation~(\ref{eq6}), and is the expected
      trend based on the $M_{\rm bh}$--$|\phi|$ relation given in
      equation~(\ref{eq1a}) and the $M_{\rm bh}$--$M_{\rm *,galaxy}$ relation
      given in equation~(\ref{eq3}), defined by galaxies with directly
      measured black hole masses (which excludes LEDA~87300, denoted by the
      open square).  The dashed grey line is given by equation~\ref{eq6B} and
      was obtained by combining equation~\ref{eq1a} and equation~\ref{eq3B}.}
    \label{Fig5}
\end{figure}

\begin{figure}
	\includegraphics[trim=1cm 2cm 2cm 1.5cm, width=\columnwidth]{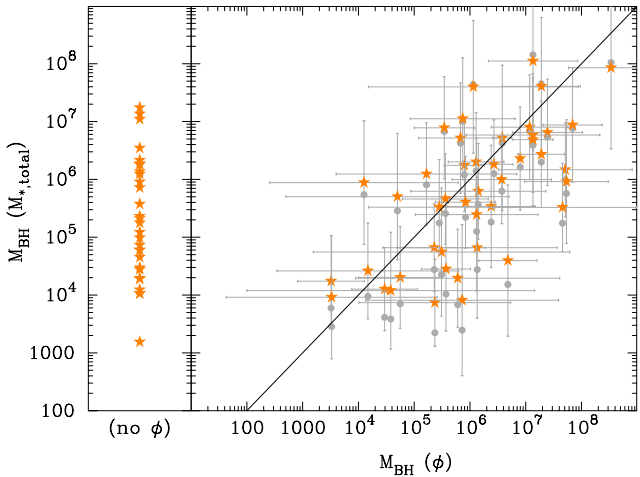}
    \caption{The orange stars show the predicted black hole masses in
      our Virgo galaxy sample, derived from both the galaxy's stellar mass
      $M_{\rm *,galaxy}$ (using equation~(\ref{eq3})) and, when available, the
      spiral arm pitch angle $|\phi|$ (using equation~(\ref{eq1a})).  The grey
      circles show the ($M_{\rm *,total}$)-based black hole masses derived
      using equation~\ref{eq3B}.
}
    \label{Fig6}
\end{figure}

\subsection{Pitch Angle vs Stellar Mass}

Combining equations~(\ref{eq1a}) and (\ref{eq3}) to eliminate $M_{\rm bh}$, one
obtains the relation
\begin{equation}
\log ( M_{\rm *,galaxy} / 6.37\times10^{10}\, M_{\odot} ) = 0.874 - 0.0645 |\phi|, 
\label{eq6}
\end{equation}
which is shown by the solid line in Figure~(\ref{Fig5}). This is, once again, the
expected relation for spiral galaxies with directly measured black hole masses. 
The trend seen here bears a resemblance to the distribution of spiral galaxies
in the diagram of pitch angle versus $B$-band absolute magnitude shown by Kennicutt
(1981, his figure~9). 

Figure~(\ref{Fig6}) displays the result of using equations~(\ref{eq1a}) and
(\ref{eq3}) to predict the black hole masses in the Virgo spiral galaxies.
This time there are many galaxies of interest in regard to potentially
harbouring an IMBH.  These findings have briefly been summarised
in Table~\ref{Tab_IMBH}.  Results for all galaxies are shown in the Appendix. 

Finally, use of equation~\ref{eq3B}, rather than equation~\ref{eq3}, results
in the relation
\begin{equation}
\log ( M_{\rm *,galaxy} / 6.37\times10^{10}\, M_{\odot} ) = 0.762 - 0.0561
|\phi|. 
\label{eq6B}
\end{equation}
It is represented by the dashed grey line in Figure~\ref{Fig5}.

\section{IMBH targets of interest}
\label{Sec_X}

\begin{table}
\centering
\caption{33 spiral galaxies with a potential IMBH}\label{Tab_IMBH}
\begin{tabular}{lccc}
\hline
Galaxy & $M_{\rm bh}$ ($M_{\rm *,total}$) & $M_{\rm bh}$ ($\phi$) & $M_{\rm bh}$ ($\sigma$) \\
       &    $M_{\odot}$                   &     $M_{\odot}$       &     $M_{\odot}$    \\
\hline
\multicolumn{4}{c}{3 estimates $< 10^5\,M_{\odot}$} \\
N4178  &   3$\times10^4$                  &   2$\times10^4$       &   1$\times10^3$ \\
N4713  &   9$\times10^3$                  &   3$\times10^3$       &   6$\times10^2$ \\

\multicolumn{4}{c}{2 estimates $< 10^5\,M_{\odot}$, no estimate $> 10^5\,M_{\odot}$} \\
IC3392  &  2$\times10^4$                  &   6$\times10^4$       &       ...   \\
N4294   &  2$\times10^4$                  &   3$\times10^3$       &       ...   \\
N4413   &  1$\times10^4$                  &   3$\times10^4$       &       ...    \\

\multicolumn{4}{c}{2 estimates $< 10^5\,M_{\odot}$, 1 estimate $\ge 10^6\,M_{\odot}$} \\
N4424   &  4$\times10^4$                  &   5$\times10^6$      &    1$\times10^5$  \\
N4470   &  1$\times10^4$                  &   4$\times10^4$      &    1$\times10^6$  \\

\multicolumn{4}{c}{1 estimate $\lesssim 10^5\,M_{\odot}$, no estimate $> 10^6\,M_{\odot}$} \\
N4197   &  7$\times10^4$                  &   2$\times10^5$       &       ...   \\
N4237   &  5$\times10^5$                  &   5$\times10^4$       &    2$\times10^5$ \\
N4298   &  5$\times10^5$                  &   4$\times10^5$       &    2$\times10^4$ \\
N4299   &  7$\times10^3$                  &   2$\times10^5$       &       ...   \\
N4312   &  1$\times10^4$                  &    ...                &    3$\times10^5$  \\
N4313   &  2$\times10^5$                  &    ...                &    2$\times10^5$   \\
N4390   &  8$\times10^3$                  &   7$\times10^5$       &       ...   \\ 
N4411b  &  3$\times10^4$                  &   4$\times10^5$       &       ...    \\
N4416   &  9$\times10^5$                  &   1$\times10^4$       &       ...    \\ 
N4498   &  2$\times10^4$                  &   6$\times10^5$       &       ...    \\ 
N4519   &  6$\times10^4$                  &   3$\times10^5$       &       ...    \\
N4647   &  4$\times10^5$                  &   8$\times10^5$       &    4$\times10^4$    \\
N4689   &  3$\times10^5$                  &   3$\times10^5$       &    2$\times10^4$    \\

\multicolumn{4}{c}{1 estimate $\lesssim 10^5\,M_{\odot}$} \\
IC3322  &  1$\times10^4$    &    ...      &     ...       \\
N4206   &  5$\times10^4$    &    ...      &      ...   \\
  N4222   &  7$\times10^4$    &    ...      &      ...   \\
N4330   &  6$\times10^4$    &    ...      &      ...     \\
  N4356   &  $\times10^5$    &    ...      &      ...     \\  
N4396   &  2$\times10^3$    &    ...      &      ...    \\ 
N4405   &  6$\times10^4$    &    ...      &      ...     \\
N4445   &  2$\times10^4$    &    ...      &      ...     \\ 
  N4451   &  1$\times10^5$    &    ...      &      ...     \\
N4522   &  2$\times10^4$    &    ...      &      ...     \\ 
N4532   &  1$\times10^4$    &    ...      &      ...     \\ 
N4606   &  3$\times10^4$    &    ...      &      ...     \\ 
N4607   &  3$\times10^4$    &    ...      &      ...     \\

\hline
\end{tabular}

Uncertainties can reach an order of magnitude, as shown 
in the Appendix Table~\ref{Tab_App1} and Figures~\ref{Fig4}, \ref{Fig1} and
\ref{Fig6}. 
\end{table}

From the previous section, we can identify five primary targets of interest:
NGC~4178 and NGC~4713 (with three black hole mass estimates less than
$10^5\,M_{\odot}$), and IC~3392, NGC~4294 and NGC~4413 (with two black hole
mass estimates less than $10^5\,M_{\odot}$ but no velocity dispersion to
provide a third black hole mass estimate).  Table~\ref{Tab_IMBH} lists these 5
galaxies along with an additional 28 galaxies which may have a central black
hole mass of less than $10^5$--$10^6\,M_{\odot}$.

The next step is to determine which of the candidate IMBH hosts harbours a
point-like nuclear X-ray source, which is likely evidence of an accreting
nuclear black hole.  We use {\it Chandra} data as the primary resource for our
search, because {\it Chandra} is the only X-ray telescope that can provide
accurate sub-arcsecond localisations of faint point-like sources (down to
$\lesssim$10 counts), thanks to the low instrumental background of its
Advanced CCD Imaging Spectrometer (ACIS). In the absence of {\it Chandra}
data, we inspected archival {\it XMM-Newton} European Photon Imaging Camera
(EPIC) data, particularly in cases when a long EPIC exposure partly made up
for the much lower spatial resolution and much higher instrumental and
background noise.  For the five primary targets identified above, two
(NGC~4178 and NGC~4713) have archival {\it Chandra}/ACIS X-ray data already
available, and one (NGC~4294) has {\it XMM-Newton}/EPIC data. The other two
(IC~3392 and NGC~4413) have recently been observed as part of our ongoing {\it
  Chandra} survey of the Virgo cluster; the results of the new observations
will be presented in a separate paper.

We re-processed and analysed the archival {\it Chandra} X-ray data using the
Chandra Interactive Analysis of Observations ({\small{CIAO}}) Version 4.9
software package (Fruscione et al.\ 2006).  For sources with a sufficient
number of counts, we extracted spectra and built response and auxiliary
response files with the {\small {CIAO}} task {\it {specextract}}, and fitted
the spectra with {\small XSPEC} version 12.9.1 (Arnaud 1996). For sources with
fewer counts, we converted between count rates and fluxes using the Portable,
Interactive Multi-Mission Simulator ({\small{PIMMS}}) software Version 4.8e,
available online\footnote{http://cxc.harvard.edu/toolkit/pimms.jsp} within the
{\it Chandra X-ray Observatory} Proposal Planning Toolkit. X-ray contour
plots, aperture photometry, and other imaging analysis was done with the
{\small {DS9}} visualization tool, part of NASA's High Energy Astrophysics
Science Archive Research Center (HEASARC) software.  For the archival {\it
  XMM-Newton} data, we used standard pipeline products (event files, images,
and source lists), downloaded from the HEASARC archive; we also used {\sc ds9}
for aperture photometry and {\sc pimms} for flux conversions.

\begin{figure}
  \includegraphics[angle=90,trim=1cm 1.7cm 0cm 1.7cm,width=1.0\columnwidth]{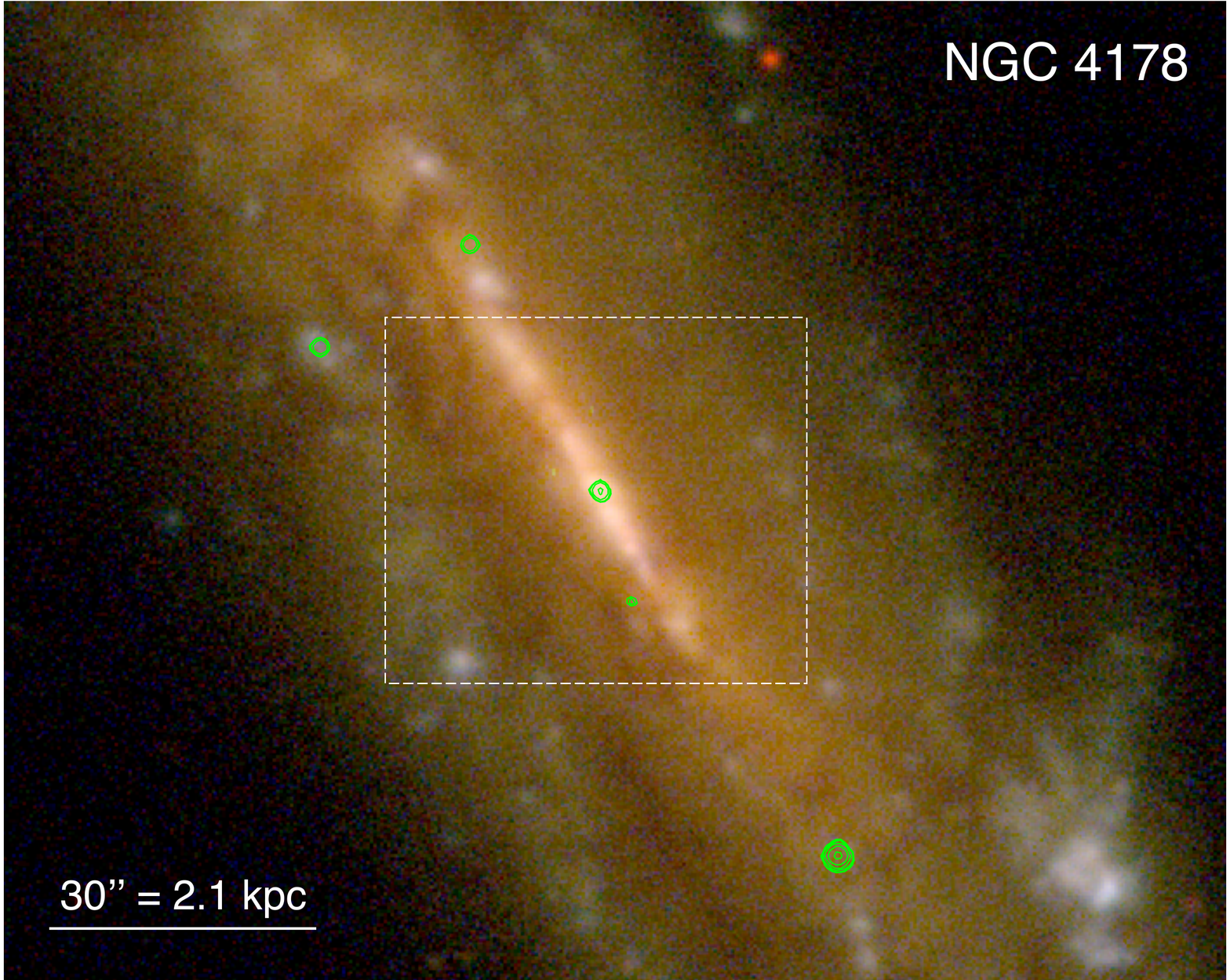}\\
  \includegraphics[angle=0,trim=0cm 7cm 0cm 5cm,width=1.0\columnwidth]{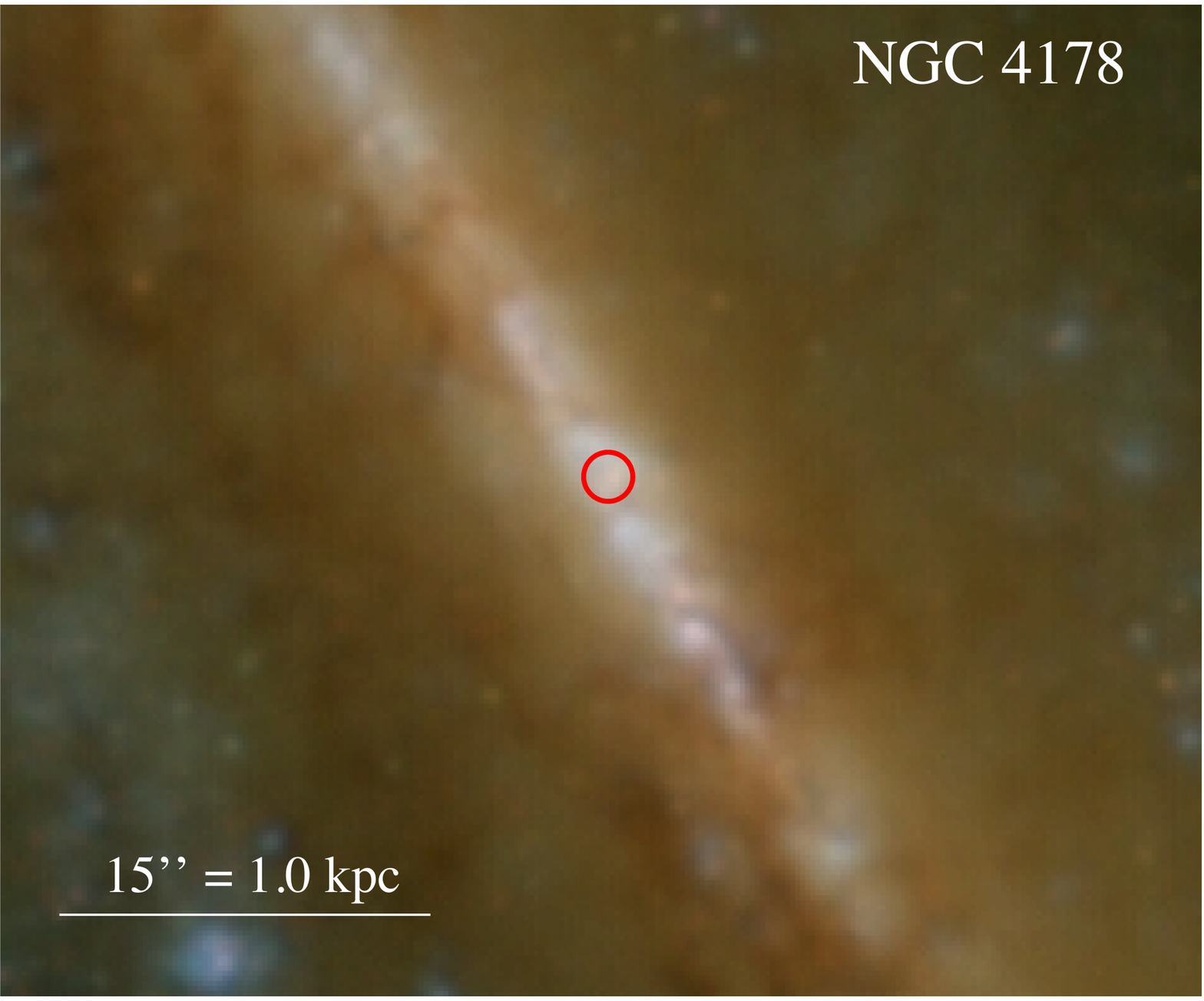}
  \caption{Top panel: SDSS image of NGC\,4178 (red = $i$ filter; green = $g$;
    blue = $u$), with {\it Chandra}/ACIS-S contours (0.3--7.0 keV band)
    overlaid in green. North is up, east is to the left. Bottom panel:
    zoomed-in view of the nuclear region, from the Next Generation
    Virgo-cluster Survey, with the position of the {\it Chandra} nuclear
    source overlaid as a red circle (radius 1$\arcsec$).}
\label{Fig7} 
\end{figure}

\begin{figure}
\hspace{-0.2cm}
  \includegraphics[angle=270,trim=2.7cm 2cm 1cm 4cm, width=1.0\columnwidth]{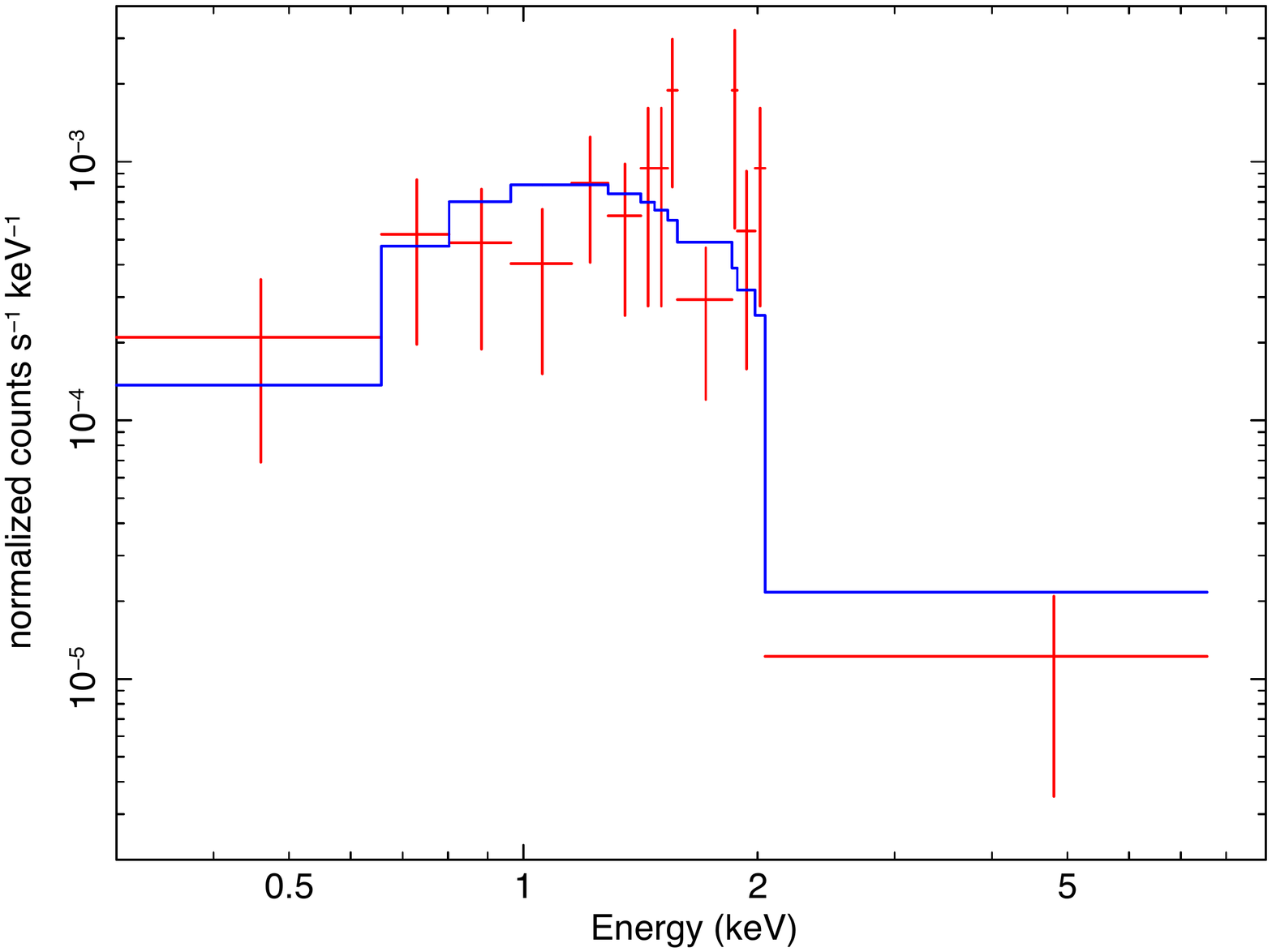}
  \caption{{\it Chandra}/ACIS-S spectrum of the nuclear source in NGC\,4178,
    fitted with a disk-blackbody model. The datapoints have been grouped to a
    signal-to-noise $>$1.5 for plotting purposes only. The fit was done on the
    individual counts, using Cash statistics. See Section 4.1 for the fit
    parameters. The sharp drop of detected counts above 2 keV disfavours a
    power-law model.}
\label{Fig8} 
\end{figure}

\subsection{NGC~4178} 

From the previous sections, we have three predictions for the black hole mass
in NGC~4178, and they all point towards a black hole in the mass range
$10^3$--$10^4\,M_{\odot}$.  This galaxy was observed by {\it Chandra}/ACIS-S
for 36 ks on 2011 February 19 (Cycle 12).  We downloaded the data from the
public {\it Chandra} archives, and reprocessed them with the {\small {CIAO}}
task {\it chandra\_repro}. We confirm the detection of an X-ray source
consistent with both the dynamical centre of the galaxy (Figure~\ref{Fig7})
and a nuclear star cluster (Satyapal et al.\ 2009; Secrest et al.\ 2012,
2013).  We extracted the source counts from a circle of radius 2$\arcsec$, and
the background counts from an annulus between radii of 3$\arcsec$ and
9$\arcsec$. As discussed by Secrest et al.\ (2012), this source is unusually
soft for an AGN. We measured a net count rate of $(2.8 \pm 0.9) \times
10^{-4}$ ct s$^{-1}$ in the 0.3--1.0 keV band, $(6.2 \pm 1.3) \times 10^{-4}$
ct s$^{-1}$ in the 1.0--2.0 keV band, and $(1.0 \pm 0.6) \times 10^{-4}$ ct
s$^{-1}$ in the 2.0--7.0 keV band. With only $\approx 36 \pm 6$ net counts, it
is clearly impossible to do any proper spectral fitting, and certainly any
fitting based on the $\chi^2$ statistics (which requires $\ga$15 counts per
bin). Nonetheless, we can fit the data with the Cash statistics (Cash 1979),
generally used for sources with a small number of counts, and constrain some
simple models. Power-law fitting based on the hardness ratio was carried out
and discussed in detail by Secrest et al.\ (2012). We re-fitted the spectrum
in {\small {XSPEC}}, with the Cash statistics, after rebinning to 1 count per
bin; we confirm that the power-law is steep, i.e.\ it is a soft spectrum, with
photon index $\Gamma = 3.4^{+1.7}_{-1.2}$, with an intrinsic absorbing column
density $N_{\rm H} = 5^{+5}_{-4} \times 10^{21}$ cm$^{-2}$ (C-statistics of
34.2 for 31 degrees of freedom). Such a steep slope, moderately high
absorption, and large uncertainty on both parameters make it difficult to
constrain the 0.3--10 keV unabsorbed luminosity: formally we obtain a 90\%
confidence limit of $L_{0.3-10} = 9^{+105}_{-6} \times 10^{38}$ erg s$^{-1}$,
consistent with the estimates of Secrest et al.\ (2012).

However, when we look at the individual detected energy of the few counts,
rather than simply considering the hardness ratio, we find that the power-law
model is inadequate. The decline in the number of detected counts above 2 keV
is very sharp (Figure~\ref{Fig8}), 
consistent with the Wien tail of an
optically thick thermal spectrum. We therefore fit the same spectrum with an
absorbed {\it diskbb} model: we obtain a C-statistic of $31.5/31$ (an
improvement at the 90\% confidence level, with respect to the power-law
fit). The best-fitting parameters are $N_{\rm H} = 1.5^{+3.3}_{-1.5} \times
10^{21}$ cm$^{-2}$ for the intrinsic absorption, $kT_{\rm in} =
0.56^{+0.35}_{-0.19}$ keV for the peak disk temperature, $r_{\rm in} =
94^{+212}_{-62} \, (\cos \theta)^{-1/2}$ km for the apparent inner-disk
radius, where $\theta$ is the viewing angle. The unabsorbed luminosity is
$L_{0.3-10} = 1.9^{+1.9}_{-0.7} \times 10^{38}$ erg s$^{-1}$. Luminosity,
temperature, and inner-disk radius are self-consistent for a stellar-mass
black hole in the high/soft state. The temperature is too high, and the radius
too small, for a supermassive black hole or even an IMBH. Invoking Occam's
razor, we argue that the most likely interpretation of the X-ray source at the
nuclear location of NGC\,4178 is a stellar-mass X-ray binary.

What to make, then, of the strong mid-IR emission in [Ne {\footnotesize{V}}]
(Satyapal et al.\ 2009; Secrest et al.\ 2012), which is usually a signature of
strong X-ray photoionisation and was the strongest argument in favour of a
hidden AGN in this galaxy? It is always possible to postulate a Compton-thick
AGN, powerful enough to supply the required luminosity; we simply argue that
this hypothesis is untestable with the available {\it Chandra}
data. Alternatively, the nuclear black hole may have been more active in the
recent past (producing the highly-ionised gas around it), but is currently in
a low state. The optical line ratios do not require an AGN, either: NGC\,4178
is classified as an H{\footnotesize II} nucleus (Ho et al.\ 1997; Decarli et
al.\ 2007; Secrest et al.\ 2012).

The uncertainty on the current luminosity, and indeed on the detection of
X-ray emission from the nuclear black hole, makes it impossible to constrain
its mass via fundamental-plane relations
(Merloni et al.\ 2003; Plotkin et al.\ 2012; Miller-Jones et al.\ 2012). 
For these relations, Secrest et al.\ (2013)
assumed an intrinsic 0.5--10 keV X-ray luminosity
$\approx$10$^{40}$ erg s$^{-1}$, a bolometric correction factor $\kappa \sim
10^3$, and an upper limit of 84.9 $\mu$Jy for the 5-GHz flux density; from
those values, they predicted a black hole mass $<$8.4 $\times 10^4 \,
M_{\odot}$. Instead, we argue that the X-ray luminosity is pure guesswork,
with no empirical constraint.
Moreover, if the nuclear black hole was indeed an IMBH, the bolometric
correction should be much lower than $10^3$; more likely, $\kappa \la 10$,
assuming a peak disk temperature $kT \ga 0.1$ keV for a black hole mass
$\la$10$^5$\,$M_{\odot}$. In summary, NGC\,4178 may host an IMBH but neither
the X-ray spectrum of the nuclear source, nor the fundamental plane relations
can be used to support this hypothesis.

Alternatively, Secrest et al.\ (2012) note that the black hole mass may be $\sim$0.1
to 1 times the mass of the nuclear star cluster ($M_{\rm nc} \sim 5\times
10^5\,M_{\odot}$: Satyapal et al.\ 2009) in this galaxy. This implies $M_{\rm
  bh} \sim (0.5-5)\times 10^5\,M_{\odot}$.  We are able to offer an
alternative prediction of the black hole mass by using the optimal $M_{\rm
  bh}$--$M_{\rm nc}$ relation extracted from Graham (2016b), which is such that
\begin{eqnarray}
&&\log(M_{\rm nc}/M_{\odot}) = \\ \nonumber
&&(0.40\pm0.13)\times\log(M_{\rm bh}/[10^{7.89}\,M_{\odot}]) + (7.64\pm0.25). 
\end{eqnarray}
From this, we derive $\log (M_{\rm bh}/M_{\odot}) = 3.04$ dex.  We conclude
that this relation offers the best additional support, beyond the initial set
of three scaling relations used in the previous section, for an IMBH in
NGC~4178.  If it has a typical Eddington ratio of say $10^{-6}$, then we
should not expect to detect it in X-rays (see Paper~I). 

\begin{figure}
  \includegraphics[angle=90,trim=1.4cm 1.6cm 0.0cm 1.6cm,width=1.0\columnwidth]{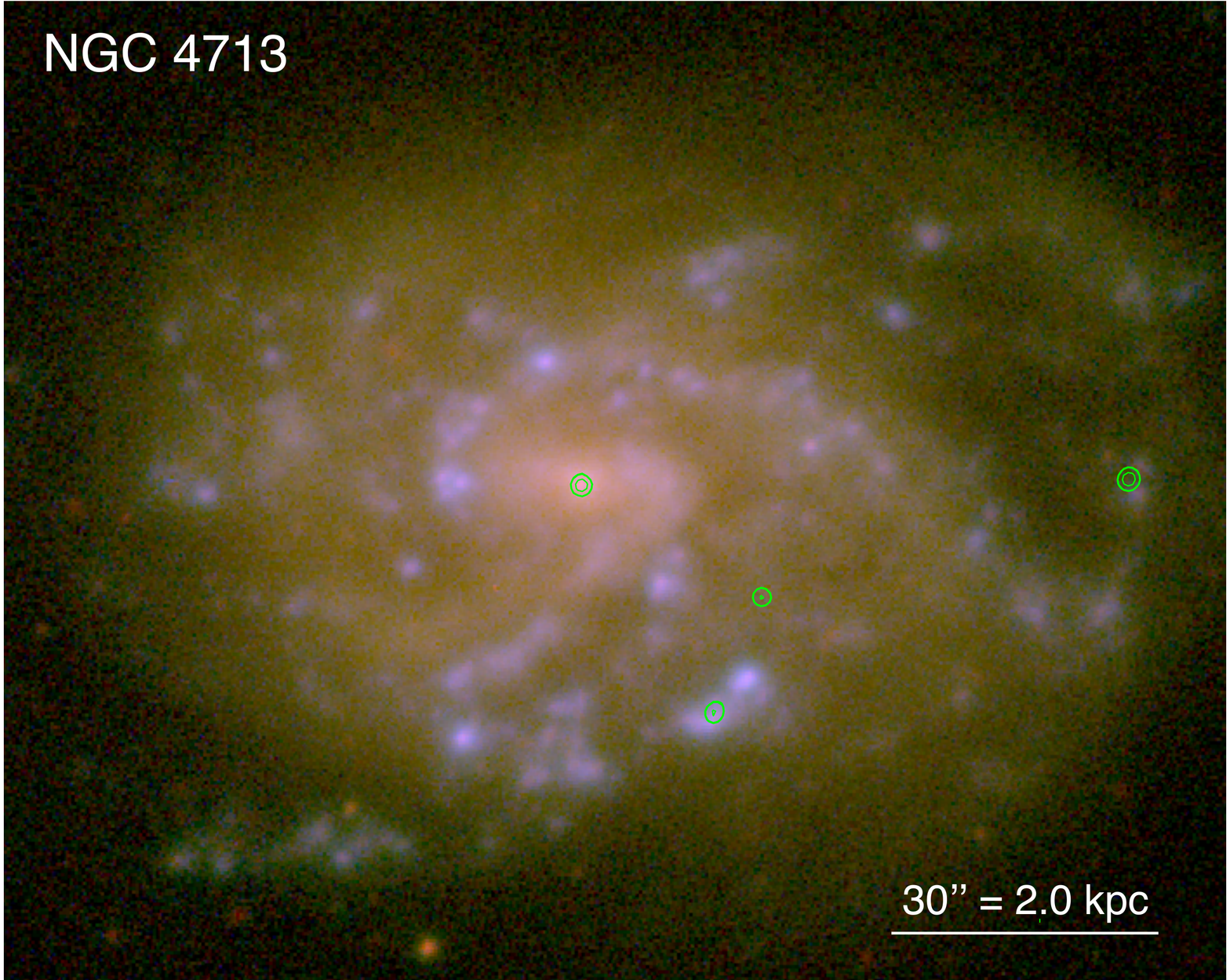}
  \caption{SDSS image of NGC\,4713, with {\it Chandra}/ACIS-S contours
    overlaid in green. North is up, east is to the left.}
\label{Fig9} 
\end{figure}

\subsection{NGC~4713} 

For NGC~4713 (SDSS J124957.86$+$051841.0), we again have three predictions for
the black hole mass, spanning (0.6--9)$\times10^3)\,M_{\odot}$. 

Based on a 4.9-ks {\it Chandra}/ACIS-S observation taken on 2003 January 28
(Cycle 4), Dudik et al.\ (2005) reported on the lack of a nuclear X-ray source
in this dwarf galaxy, with a 0.3--10 keV upper limit of $\approx$4$ \times
10^{-4}$ ct s$^{-1}$; looking in the mid-IR, Satyapal et al.\ (2009) also
excluded an active nucleus.  However, Nagar et al.\ (2002) had included it in
their list of BPT (Baldwin et al.\ 1981) `composite galaxies', as did
Reines, Greene \& Geha (2013).  Furthermore, Terashima et al.\ (2015) found an
X-ray source at the nuclear position in a 52.2-ks {\it XMM-Newton} European
Photon Imaging Camera observation, part of the {\it XMM-Newton} Serendipitous
Source Catalog Data Release 3 (Watson et al.\ 2009). The unabsorbed 0.3--10
keV flux is reported as $\approx 5\times 10^{-14}$ erg cm$^{-2}$ s$^{-1}$,
with the addition of a `weak hint' of an Fe-K line at 6.4 keV. Based on the
ratios between the FIR luminosities at 18 $\mu$m and 90 $\mu$m (from the AKARI
survey: Kawada et al.\ 2007; Ishihara et al.\ 2010), and the X-ray luminosity,
Terashima et al.\ (2015) classified the nucleus of NGC\,4713 as an unobscured
transition object between LINERS and H{\footnotesize {II}} nuclei. The lower
spatial resolution of {\it XMM-Newton} makes it impossible to determine
whether the faint nuclear X-ray emission is point-like, from an AGN, or
extended, from hot gas in a star-forming region.

We reprocessed and re-examined the 4.9-ks {\it Chandra}/ACIS-S observation. In
contrast to the conclusions of Dudik et al.\ (2005), we do find a point-like X-ray
nucleus (see Figure~\ref{Fig9}), 
located within $0\arcsec.2$ of the optical nucleus as defined by
SDSS-DR12 (Alam et al.\ 2015) and {\it Gaia} Data Release 2 
(Gaia Collaboration et al.\ 2018). 
We measure 
a net count rate in the 0.3--7.0 keV band of $2.0^{+1.3}_{-0.9} \times
10^{-3}$ ct s$^{-1}$, i.e.\ 10 raw counts and 0.2 background counts. The
errors reported here are 90\% confidence limits calculated from the Tables of
Kraft et al.\ (1991), suitable for sources with a low number of counts. Source counts 
are detected in all three standard bands (soft, 0.3--1 keV; medium, 1--2 keV; hard, 2--7 keV), 
which is consistent with a power-law spectrum. Assuming a
power-law spectrum with photon index $\Gamma = 1.7$, and line-of-sight column
density $N_{\rm H} = 2 \times 10^{20}$ cm$^{-2}$, we find with {\sc pimms} that 
the net count rate
corresponds to a 0.3--10 keV unabsorbed flux of $(1.4^{+0.9}_{-0.6}) \times
10^{-14}$ erg cm$^{-2}$ s$^{-1}$ (slightly lower than the {\it XMM-Newton}
flux, which may include a hot gas contribution). At the distance of 13.2 Mpc
for NGC~4713, this implies a luminosity $L_{0.3-10} = 3.0^{+1.9}_{-1.4}
\times 10^{38}$ erg s$^{-1}$. We also obtained an essentially identical 
estimate of $L_{0.5-7} \approx 3.1 \times 10^{38}$ erg s$^{-1}$ using the 
{\sc ciao} task {\it srcflux} within {\sc ds9}, with the same input spectral model 
and column density.

\subsection{Galaxies with 2 BH mass estimates $< 10^5 M_{\odot}$}

NGC~4294 was observed by {\it XMM-Newton}/EPIC on 2016 June 10, for 46 ks.  We
found no sources at the nuclear position, in the stacked EPIC pn and MOS
image.  The nearest point source is located $\approx$9$\arcsec$ from the optical
nuclear position, way beyond the possible astrometric uncertainty of the EPIC
image. We estimate a 90\% confidence limit for the 0.3--10 keV nuclear
luminosity of $L_{0.3-10} < 1 \times 10^{38}$ erg s$^{-1}$, for a distance of
17.0 Mpc.

Among the other four galaxies with two black hole mass estimates $< 10^5
M_{\odot}$ (Table~\ref{Tab_IMBH}), three (IC~3392, NGC~4413, and NGC~4424)
have been observed by {\it Chandra} this year for our Virgo survey, and we
will present the results in a separate paper. The fourth galaxy, NGC~4470, was
observed by {\it Chandra} several times between 2010 and 2016: twice with
ACIS-I, and four times with ACIS-S. However, in all cases, it was not the
primary target of the observation, and was located several arcmin away from
the aimpoint, with a resulting degradation of the point spread function at its
location.  Moreover, in three of the four ACIS-S observations, the nuclear
position of NGC~4470 fell onto the less sensitive S2 chip, rather than the S3
chip. Of all the available datasets, the most useful one for our investigation
is from a 20-ks observation taken on 2010 November 20 (Cycle 12): it is the
only observation in which NGC~4470 is on the S3 chip, only $\approx$4$'$ from
the aimpoint.  From this observation, we found excess emission centred at
$\approx$1$\arcsec$ of the SDSS and {\it Gaia} optical nuclear positions
(smaller than the positional uncertainty of the X-ray source at that off-axis
position and for the small observed number of counts), with a net count rate
of $(5 \pm 2) \times 10^{-4}$ ct s$^{-1}$ ({\it i.e.}, $\approx$ 10 net
counts) in the 0.3--7.0 keV band. For an absorbing column density $N_{\rm H} =
1.7 \times 10^{20}$ cm$^{-2}$ a power-law photon index $\Gamma = 1.7$, and a
distance of 18.8 Mpc, this corresponds to an unabsorbed luminosity $L_{0.3-10}
= 2.2^{+1.9}_{-1.2} \times 10^{38}$ erg s$^{-1}$.

\subsection{Galaxies with 1 BH mass estimate $\lesssim 10^5 M_{\odot}$}

\begin{figure}
  \includegraphics[angle=90,trim=1.4cm 1.6cm 0.0cm 1.6cm,width=1.0\columnwidth]{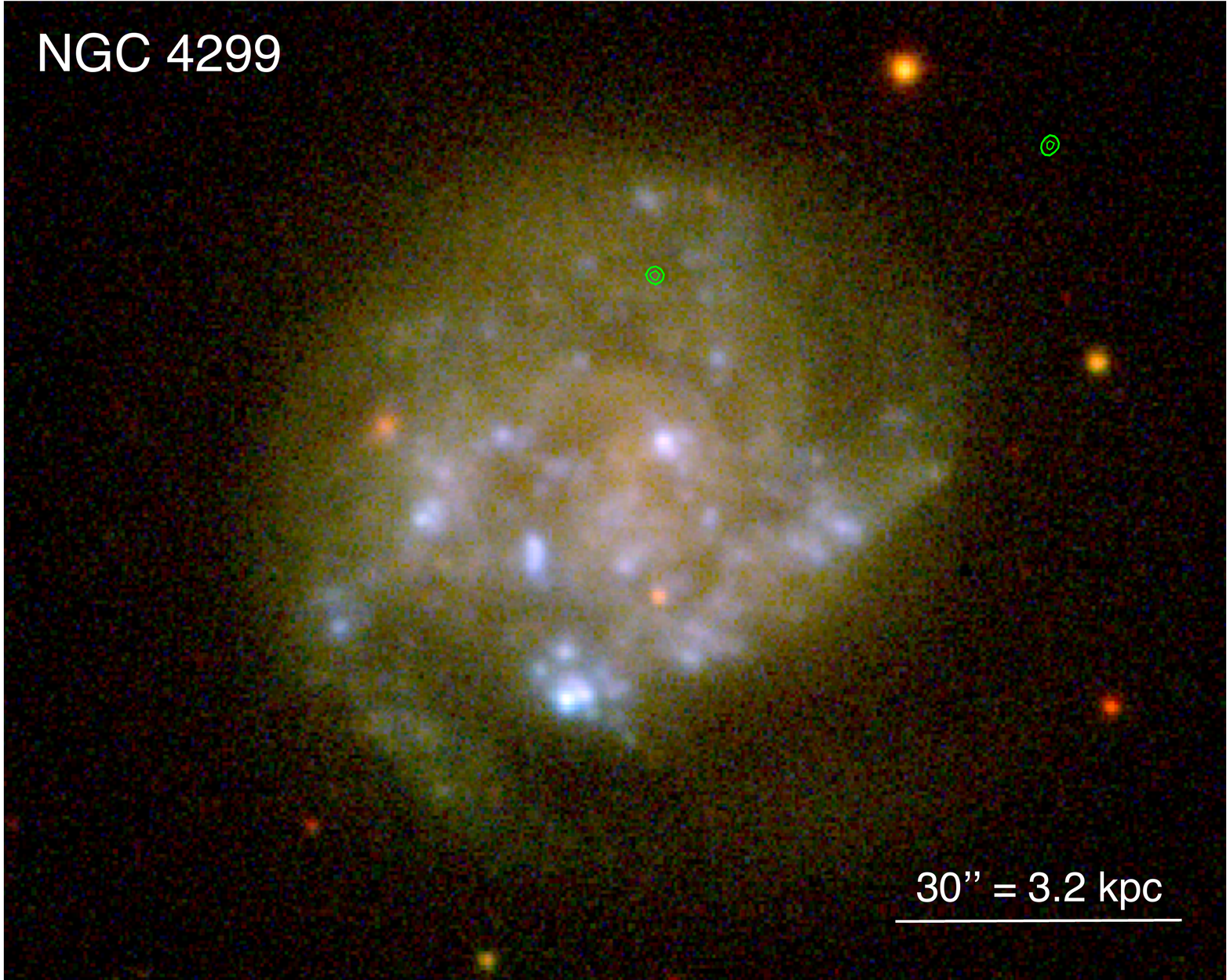}
  \caption{SDSS image of NGC~4299, with {\it Chandra}/ACIS-S contours overlaid
    in green (seen to the right of the galaxy). North is up, east is to the left.}
\label{Fig4299} 
\end{figure}

Finally, we examined the 26 galaxies with one black hole mass estimate
$\lesssim 10^5 M_{\odot}$ (Table~\ref{Tab_IMBH}). Twenty of them have been
observed as part of our Virgo survey, and we will report on them
elsewhere. The other three, NGC~4299, NGC~4647 and NGC~4689, already had
archival {\it Chandra} data.

NGC~4299 was observed by {\it Chandra}/ACIS-S on 2007 November 18, for 5.0 ks. 
We do not find any significant emission
at the nuclear position (see Figure~\ref{Fig4299}); 
we place a 90\% upper limit of $8 \times 10^{-4}$ count 
s$^{-1}$ in the 0.3--7 keV band, using the tables of Kraft et al.\ (1991). Assuming
line-of-sight Galactic absorption $N_{\rm H} = 2.5 \times 10^{20}$ cm$^{-2}$,
and a power-law spectrum with photon index $\Gamma = 1.7$, this corresponds to
a luminosity $L_{0.3-10} < 7 \times 10^{38}$ erg s$^{-1}$ at the distance of 21.9 Mpc.

NGC~4647 was observed by {\it Chandra}/ACIS-S on six visits between 2000 and
2011 (2000 April 20: 38 ks; 2007 January 30: 52 ks; 2007 February 01: 18 ks;
2011 August 08: 85 ks; 2011 August 12: 14 ks; 2011 February 24: 101 ks), for a
total of $\approx$308 ks. On all occasions, the aimpoint was located at the
main target of the observation, the nearby E galaxy NGC~4649; however,
NGC~4647 is only $\approx$2$'$.5 away, and falls within the S3 chip in all six
datasets. We inspected each observation individually, and we then used the
{\sc ciao} script {\it merge\_obs} to created a stacked, exposure-corrected
image. We used point-like optical/X-ray associations to align the {\it
  Chandra} image onto the SDSS optical image, so that the two frames coincide
within $\lesssim$0$\arcsec$.3. There is no point-like X-ray source at the
nuclear location, defined by SDSS (also in agreement with {\it Gaia}'s nuclear
position within the same uncertainty). The nearest point-like X-ray source
(most likely an X-ray binary) is located $\approx$3$\arcsec$ to the west of
the nuclear location, well above its positional uncertainty. The 90 percent
upper limit of the (undetected) nuclear source is $3.6 \times 10^{-5}$ ct
s$^{-1}$ in the 0.3--7 keV band. To convert this upper limit into a luminosity
limit, we took a weighted average of the contributions over the different
observation cycles, to take into account the change in the response of the
ACIS detector; we also assumed, as usual, a line-of-sight Galactic absorption
(in this case, $N_{\rm H} = 2.1 \times 10^{20}$ cm$^{-2}$) and a power-law
spectrum with photon index $\Gamma = 1.7$. We conclude that the nuclear X-ray
luminosity is $L_{0.3-10} < 1 \times 10^{37}$ erg s$^{-1}$, at the distance of
17.6 Mpc.

NGC~4689 was observed by {\it Chandra}/ACIS-S for 5 ks on 2007 May 7 (Cycle
8). We do not detect any net emission at the nuclear position; the Bayesian
90\% confidence limit on the net count rate is $\approx$5 $\times 10^{-4}$ ct
s$^{-1}$. For a line-of-sight Galactic absorption $N_{\rm H} = 2.0 \times
10^{20}$ cm$^{-2}$ and a photon index $\Gamma = 1.7$, we obtain an upper limit
to the nuclear black hole luminosity of $L_{0.3-10} < 1.2 \times 10^{38}$ erg
s$^{-1}$, at the distance of 16.4 Mpc.

A summary of the X-ray observation exposure times, count
rates, and luminosities is provided in Table~\ref{Tab_X_Sum}.

\begin{table*}
    \centering
    \caption{Summary of the X-ray observations used for this work and of their
      respective nuclear X-ray properties.} 
    \begin{tabular}{lccccc}
    \hline
Galaxy  & Observatory & Date & Exp Time & Count Rate$^a$ & $L_{0.3-10}$ \\
       &     & & (ks) & (ct s$^{-1}$) & (erg s$^{-1}$)  \\  
    \hline
NGC~4178   & {\it Chandra} & 2011-02-19 & 36 & $\left(1.0^{+0.2}_{-0.2}\right)
\times 10^{-3}$ & $\left(1.9^{+1.9}_{-0.7}\right) \times 10^{38}$\\[3pt]
NGC~4713   & {\it Chandra} & 2003-01-28 & 4.9 &
$\left(2.0^{+1.3}_{-0.9}\right) \times 10^{-3}$ &
$\left(3.0^{+1.9}_{-1.4}\right) \times 10^{38}$\\[3pt]
NGC~4294   & {\it XMM-Newton} & 2006-06-10 & 46 & $< 1 \times 10^{-3}$ & $< 1
\times 10^{38}$\\[3pt]
NGC~4470   & {\it Chandra} & 2010-11-20 & 20 & $\left(0.5^{+0.2}_{-0.2}\right)
\times 10^{-3}$ & $\left(2.2^{+1.9}_{-1.2}\right) \times 10^{38}$\\[3pt]
NGC~4299   & {\it Chandra} & 2007-11-18 & 5.0 & $< 0.8 \times 10^{-3}$ & $< 7
\times 10^{38}$\\[3pt]
NGC~4647   & {\it Chandra} & 2000-04-20 &  38 & & \\
           &   &             2007-01-30 &  52  & & \\ 
           &    &            2007-02-01  &  18 & & \\ 
           &        &        2011-02-24 & 101 & & \\ 
           &     &           2011-08-08 &  85  & & \\
           &      &          2011-08-12  &  14  & & \\ 
           &          &      (stacked)    &  308 & $< 0.036 \times 10^{-3}$ &
$< 0.1 \times 10^{38}$\\[3pt]
NGC~4689   & {\it Chandra} & 2007-05-07 & 5.0 & $< 0.5 \times 10^{-3}$ & $< 1.2 \times 10^{38}$\\[3pt]
    \hline
    \end{tabular}

$^a$ For {\it Chandra} observations: observed ACIS-S count rate in the 0.3--7
    keV band; for {\it XMM-Newton}: observed EPIC-pn count rate in the 0.3--10
    keV band.
    \label{Tab_X_Sum}
\end{table*}

\section{Discussion}
\label{Sec_Disc}

Observational evidence for $\sim$12, 30 and 60 solar mass black holes already exists
(Reid et al.\ 2014; Abbott et al.\ 2016, 2017). It is also understood that 
in today's universe, the end product of a massive star will be a `stellar 
mass' black hole less than 80--100 $M_{\odot}$, but 
near this limit if the star's metallicity was low ($10^{-2}$--$10^{-3}$ solar) and
the mass loss from its stellar wind was low (Belczynski et al.\ 2010; Spera et
al.\ 2015; Spera \& Mapelli 2017).  
While some authors have advocated that the seed masses which gave rise to
the supermassive black holes at the centres of galaxies 
started out with masses of $10^5$--$10^6\,
M_{\odot}$ (e.g.\ Turner 1991; Loeb \& Rasio 1994), and many AGN are known to have 
 $10^5$--$10^6$ $M_{\odot}$ black holes (e.g.\ Graham \& Scott 2015, and references therein), 
the assumption that there are not black holes with masses of $10^2$--$10^5$ $M_{\odot}$ 
need not hold.  The perceived need for massive seeds was originally invoked because, 
under the assumption of spherical accretion, there was not sufficient time to
grow the massive quasars observed in the early Universe.  However,
it is possible to grow black holes at a much faster rate than the idealised
and restrictive Eddington accretion rate (e.g.\ Alexander 
\& Natarajan 2014; Nayakshin et al.\ 2012). 
Moreover, even if the {\it massive} quasars did form from massive seeds
(Pacucci et al.\ 2016), 
there may still be a continuum of BH masses, 
perhaps with today's IMBHs born from Pop III and II.5 stars, 
or from other processes, as noted in Section~\ref{SecIntro}. 
There are, therefore, reasons to expect that IMBHs with $10^2 < M_{\rm bh}/M_{\odot} 
< 10^5$ should exist.

Just as initial searches for exoplanets found the larger ones first, and
surveys of galaxies found the bright Hubble-Jeans sequence (Jeans 1919, 1928;
Hubble 1926, 1936) prior to the detection of low surface brightness galaxies,
sample selection effects hinder the detection of IMBHs in galactic nuclei.
Their gravitational spheres-of-influence are too small to be spatially
resolved with our current instrumentation.  Furthermore, ambiguity also arises
because the energy levels of their current low-accretion activity overlaps
with that of highly-accreting stellar mass black holes in X-ray
binaries. There is, however, no obvious physical reason why these IMBHs should
not exist, and a small number of candidates are known, including the
already-mentioned $\sim$10$^4 \, M_{\odot}$ black hole near/inside ESO~243-49
(Farrell et al.\ 2009; Yan et al.\ 2015; Webb et al.\ 2017; Soria et
al.\ 2017), plus the nuclear black holes in LEDA~87300 (Baldassare et
al.\ 2015; whose mass estimate was halved in Graham et al.\ 2016, see also
Baldassare et al.\ 2017) and in NGC~404 (with a 3$\sigma$ upper limit on its
black hole mass of $1.5\times 10^5\, M_{\odot}$: Nguyen et al.\ 2017).

There is also a series of studies (Pardo et al.\ 2016; Mezcua et al.\ 2016,
2018) which have estimated the masses of distant low mass black holes using a
near-linear $M_{\rm bh}$--$M_{\rm *tot}$ relation for AGN from Reines \&
Volonteri (2015).  However, it should be noted that Reines \& Volonteri (2015)
appear unaware of, or reject, the bend in the $M_{\rm bh}$--$M_{\rm
  *spheroid}$ diagram (see Graham \& Scott 2015), and the associated bend in
the $M_{\rm bh}$--$M_{\rm *tot}$ diagram which is evident in the data they
present.  Fitting a log-linear relation to galaxies that they consider to
contain a classical bulge rather than a pseudobulge, the right hand panel of
figure~10 in Reines \& Volonteri (2015) reveals that all galaxies with $M_{\rm
  bh} \lesssim 10^8\,M_{\odot}$ (except for the stripped compact elliptical
galaxy M~32 which should be down-weighted in this diagram due to its rare
nature relative to normal galaxies) reside below their $M_{\rm bh}$--$M_{\rm
  *tot}$ relation for classical bulges and elliptical galaxies.  Many more
galaxies with directly measured black hole masses also reside below their
relation, but they were labelled ``pseudobulges'' and excluded by Reines \&
Volonteri (2015).  Given that the bulge-to-total mass ratio tends to decrease
as one progresses to lower mass spiral galaxies, the $M_{\rm bh}$--$M_{\rm
  *tot}$ relation for spiral galaxies (Davis et al.\ 2018b) is steeper than
the near-quadratic relation for the bulges of spiral galaxies (e.g.\ Scott et
al.\ 2013; Savorgan et al.\ 2016; Davis et al.\ 2018a).  Reines \& Volonteri
(2015) instead suggest that their AGN sample --- used to define their
near-linear $M_{\rm bh}$--$M_{\rm *tot}$ relation for AGN --- reside in
pseudobulges that have an $M_{\rm bh}/M_{\rm *,tot}$ ratio of $\approx$0.03
percent at $M_{\rm *tot} = 10^{11}\,M_{\odot}$.  However, their distribution
of AGN data does not match the distribution of spiral galaxies (also alleged
to contain pseudobulges) with directly measured black hole masses and stellar
masses derived from space-based infrared images. As such, while the dwarf
galaxies studied by Pardo et al.\ (2016) and Mezcua et al.\ (2016, 2018) may
contain IMBHs, it may be worthwhile revisiting their black hole masses.

A growing number of alleged and potential IMBHs, not located at the center of
their host galaxy, are also known (e.g.\ Colbert \& Mushotzky 1999; Farrell et
al.\ 2009, 2014; Soria et al.\ 2010; Webb et al.\ 2010, 2014; Liu et
al.\ 2012; Secrest et al.\ 2012; Sutton et al.\ 2012; Kaaret \& Feng 2013;
Miller et al.\ 2013; Cseh et al.\ 2015; Mezcua et al.\ 2015; Oka et al.\ 2016;
Pasham et al.\ 2014, 2015).  It has been theorised that some of these may have
previously resided at the centre of a galaxy: perhaps from a stripped
satellite galaxy or minor merger (e.g.\ Drinkwater et al.\ 2003), or perhaps
they were dynamically ejected from the core of their host galaxy
(e.g.\ Merritt et al.\ 2009).  This latter phenomenon may occur due to the
gravitational recoiling of a merged black hole pair (e.g.\ Bekenstein 1973;
Favata et al.\ 2004; Herrmann et al.\ 2007; Nagar 2013).  Alternatively, or
additionally, IMBHs may have formed in their off-centre location.  Such
speculation should, however, be tempered at this point because, as noted in
Section~\ref{SecIntro}, many such past IMBH candidates can be explained as
super-Eddington accretion onto stellar-mass compact objects
(Feng \& Soria 2011; Kaaret et al.\ 2017). 

There are many methods, beyond those already employed here, 
that can be used to identify, and probe the masses of, black holes. 
This includes reverberation 
mappings of AGN (e.g.\ Bahcall, Kozlovsky \& Salpeter 1972; Blandford \& McKee
1982; Netzer \& Peterson 1997), the `fundamental plane of black hole
activity' (Merloni et al.\ 2003; Falcke et al.\ 2004), spectral modelling of the
high-energy X-ray photon coming from 
the hot accretion discs around IMBHs (Pringle \& Rees 1972; Narayan \& Yi
1995), high-ionization optical emission lines (Baldwin et al.\ 1981; Kewley et
al.\ 2001); and high spatial resolution observations 
of maser emission using radio and millimetre/submillimeter interferometry (e.g.\ Miyoshi et
al.\ 1995; Greenhill et al.\ 2003; Humphreys et al.\ 2016; Asada et al.\ 2017). 
In addition, 
the merging of black holes is now quite famously known to produce 
gravitational radiation during their orbital decay (Abbott et al.\ 2016).  The
merging of galaxies containing their own central IMBH is similarly expected to
result in the eventual merging of these black holes.  The Kamioka
Gravitational Wave Detector (KAGRA: Aso et al.\ 2013) will be a 3-km long
underground interferometer in Japan that is capable of detecting the
gravitational radiation emanating from collisions involving black holes with
masses up to 200 $M_{\odot}$ (T\'apai et al.\ 2015).  The planned Deci-Hertz
Interferometer Gravitational wave Observatory (DECIGO: Kawamura et al.\ 2011)
and the European, Laser Interferometer Space Antenna (LISA) Pathfinder
mission\footnote{\url{http://sci.esa.int/lisa-pathfinder/}} (Anza et
al.\ 2005; McNamara 2013), with their greater separation of mirrors, will be
able to detect longer wavelength gravitational waves, and thus better reach
into the domain of intermediate-mass and supermassive black hole mergers, the
latter of which are currently being searched for via `pulsar timing arrays'
(PTAs) (e.g.\ Hobbs et al.\ 2010; Kramer \& Champion 2013; Shannon et
al.\ 2015).  
A key constraint to the expected detection threshold of
such signals from PTAs -- in particular the background of cosmic ripples from
the merger of massive black holes (themselves arising from the merger of
galaxies) -- is the (black hole)-to-(host galaxy/bulge) mass ratio (see
equation~(\ref{eq3}) for spiral galaxies).  An additional source of long
wavelength gravitational radiation will arise from the inspiral of compact
stellar mass objects, such as neutron stars and black holes, around these
IMBHs (Mapelli et al.\ 2012). It is reasonable to expect that the densely
packed nuclear star clusters, which coexist with low-mass SMBHs (e.g.\
Gonz{\'a}lez Delgado et al.\ 2008; Seth et al.\ 2008; Graham \& Spitler 2009),
will similarly surround many IMBHs.  Gravitational radiation, and the
gravitational tidal disruption of ill-fated stars that venture too close to
these black holes (Komossa et al.\ 2009, Komossa 2013 and references therein; 
Zhong et al.\ 2015; Stone \& Metzger 2016; Lin et
al.\ 2018), are therefore expected from these astrophysical entities.
There is, therefore, an array of future observations which could yield further
confidence and insight into the realm of IMBHs. 


In the pursuit of galaxies that may harbour (some of) the largely missing
population of IMBHs, we have predicted the black hole
masses in 74 spiral galaxies in the Virgo cluster that will be imaged with
the ACIS-S detector on the {\it Chandra X-ray Observatory}.  
Previously, Gallo et al.\ (2008) performed a complementary investigation
looking at 100 early-type galaxies in the Virgo cluster.  
However, they only used two global properties of the galaxies ($\sigma$ and
$M_{\rm *,galaxy}$) to predict the black hole masses, and their predictions
differed systematically and significantly from each other (Gallo et al.\ 2008,
their figure~4), revealing that either one, or both, of 
their black hole scaling relations was in error.  
That offset, which reached 3 orders of magnitude at the low mass end, is
investigated and reconciled in Paper~I.  Here, we have used three 
global properties of spiral galaxies ($\sigma$, $M_{\rm *,galaxy}$ and spiral arm
pitch angle $\phi$) to predict the black hole masses in our spiral galaxy
sample. Moreover, our updated scaling relations are internally consistent with each
other and do not contain any dramatic systematic bias.  Table~\ref{Tab_Calib} provides a
sense of what galaxy parameter values are associated with a given set of black hole
masses.  Based on our estimates of these galaxies' stellar masses, 
33 of the 74 galaxies are predicted to have a
black hole mass less than $10^5$--$10^6\, M_{\odot}$ (see Table~\ref{Tab_IMBH}).

\begin{table}
\centering
\caption{Black hole calibration points}\label{Tab_Calib}
\begin{tabular}{cccc}
\hline
$M_{\rm bh}$  &  $M_{\rm *,total}$                      & $\phi$ & $\sigma$     \\
 $M_{\odot}$  &    $M_{\odot}$                          & [deg]  & km s$^{-1}$  \\
\hline
$10^9$    &    2.9$\times10^{11}$ (2.4$\times10^{11}$)  &   3.0  &  293    \\
$10^8$    &    1.2$\times10^{11}$ (1.1$\times10^{11}$)  &   9.1  &  195    \\
$10^7$    &    5.1$\times10^{10}$ (5.3$\times10^{10}$)  &  15.2  &  130    \\
$10^6$    &    2.1$\times10^{10}$ (2.5$\times10^{10}$)  &  21.3  &   86    \\
$10^5$    &    8.9$\times10^{9}$  (1.2$\times10^{10}$)  &  27.4  &   57    \\
$10^4$    &    3.7$\times10^{9}$  (5.5$\times10^{9}$)   &  33.5  &   38    \\
$10^3$    &    1.6$\times10^{9}$  (2.6$\times10^{9}$)   &  39.6  &   25    \\
$10^2$    &    6.6$\times10^{8}$  (1.2$\times10^{8}$)   &  45.7  &   17    \\
\hline
\end{tabular}

Reversing equations~\ref{eq3} (\ref{eq3B}), \ref{eq1b} and \ref {eq2}, we
provide the total galaxy stellar mass, spiral arm pitch angle and stellar
velocity dispersion that corresponds to the black hole masses listed in
column~1, respectively. 
\end{table}

The black hole mass estimates presented here shall be used in a number of
forthcoming papers once imaging from the new 
{\it Chandra} Cycle 18 Large Project `Spiral
galaxies of the Virgo cluster' (Proposal ID: 18620568) 
is completed. 
Given the low degree of scatter about the $M_{\rm
  bh}$--$|\phi|$ relation, it appears to be the most promising relation to use
in the search for IMBHs in late-type galaxies.  In future work, we intend to identify
those late-type spiral galaxies with open, loosely-wound spiral arms,
i.e.\ those expected to have the lowest mass black holes at their centre, and
then check for the signature of a hot accretion disk heralding the presence of
potentially further IMBHs.

\section*{Acknowledgements}

This research was supported under the Australian Research
Council's funding scheme DP17012923. 
Part of this research was conducted within the Australian Research Council's 
  Centre of Excellence for Gravitational Wave Discovery (OzGrav), through
  project number CE170100004. 
Support for this work was provided by the National Aeronautics and Space
Administration through Chandra Award Number 18620568. 
This research has made use of the NASA/IPAC Extragalactic Database (NED).
This publication makes use of data products from the Two Micron All Sky Survey.
We acknowledge use of the HyperLeda database (http://leda.univ-lyon1.fr).
This research has made use of the GOLDMine Database.

\appendix
\section{Galaxy sample and predicted black hole masses} 

\begin{table*}
\label{Tab_App1}
\centering
\caption{Predicted black hole masses} 
\resizebox{\textwidth}{!}{%
\begin{tabular}{llccccccc}
\hline
 Galaxy  &  Type       &   Dist.Mod.   & $|\phi|$ (band) &  $\log M_{\rm bh}(|\phi|)$  &   $\sigma$      &  $\log M_{\rm bh}(\sigma)$  & $\log M_{\rm *,gal}$ & $\log M_{\rm bh}(M_{\rm *,gal})$ \\
         &             &               & [deg]           &     [dex]     &  km s$^{-1}$      &   [dex]       &   [dex]        &    [dex]   \\      
\hline
IC3322   & SAB(s)cd    & 31.7$\pm$0.2  &     ...                              &    ...        &          ...     &    ...        &   9.6$\pm$0.1  & 4.0$\pm$0.7  (3.5$\pm$0.8)  \\
IC3322A  & SB(s)cd     & 31.9$\pm$0.4  &     ...                              &    ...        &          ...     &    ...        &  10.1$\pm$0.2  & 5.4$\pm$0.8  (5.1$\pm$0.9)  \\
IC3392   & SAb         & 30.7$\pm$0.5  &  28.2$\pm$ 3.9 (GALEX/FUV       )    & 4.8$\pm$0.8   &          ...     &    ...        &   9.7$\pm$0.2  & 4.3$\pm$0.9  (3.8$\pm$1.0)  \\
N4178    & SB(rs)dm    & 30.7$\pm$0.4  &  31.6$\pm$ 9.3 (Spitzer/IRAC3   )    & 4.2$\pm$1.6   &   26.0$\pm$ 3.9  & 3.1$\pm$0.9   &   9.7$\pm$0.2  & 4.4$\pm$0.8  (4.0$\pm$0.9)  \\
N4192    & SAB(s)ab    & 30.7$\pm$0.4  &     ...                              &    ...        &  129.0$\pm$19.4  & 7.0$\pm$0.7   &  10.7$\pm$0.2  & 7.0$\pm$1.0  (7.0$\pm$1.0)  \\
N4197    & Sd          & 32.2$\pm$0.3  &  24.6$\pm$ 5.0 (SDSS/g          )    & 5.4$\pm$0.9   &          ...     &    ...        &   9.9$\pm$0.1  & 4.8$\pm$0.8  (4.4$\pm$0.8)  \\
N4206    & SA(s)bc     & 31.3$\pm$0.4  &     ...                              &    ...        &          ...     &    ...        &   9.8$\pm$0.2  & 4.7$\pm$0.8  (4.2$\pm$0.9)  \\
N4212    & SAc         & 31.3$\pm$0.3  &  21.5$\pm$ 1.1 (Spitzer/IRAC2   )    & 5.9$\pm$0.4   &   61.0$\pm$ 9.2  & 5.1$\pm$0.8   &  10.4$\pm$0.1  & 6.2$\pm$0.8  (6.1$\pm$0.9)  \\
N4216    & SAB(s)b     & 31.0$\pm$0.3  &  20.5$\pm$10.8 (Spitzer/IRAC2   )    & 6.1$\pm$1.9   &  196.0$\pm$29.4  & 8.0$\pm$0.6   &  10.9$\pm$0.1  & 7.6$\pm$1.1  (7.6$\pm$1.1)  \\
N4222    & Sd          & 31.8$\pm$0.4  &     ...                              &    ...        &          ...     &    ...        &   9.9$\pm$0.2  & 4.9$\pm$0.8  (4.5$\pm$0.9)  \\
N4237    & SAB(rs)bc   & 31.4$\pm$0.8  &  28.5$\pm$11.5 (CFHT/u          )    & 4.7$\pm$2.0   &   62.0$\pm$ 9.3  & 5.2$\pm$0.8   &  10.2$\pm$0.3  & 5.7$\pm$1.1  (5.5$\pm$1.2)  \\
N4254    & SA(s)c      & 30.9$\pm$0.3  &  23.6$\pm$ 2.6 (GALEX/FUV       )    & 5.5$\pm$0.5   &   98.0$\pm$14.7  & 6.3$\pm$0.7   &  10.7$\pm$0.1  & 6.9$\pm$0.9  (6.8$\pm$0.9)  \\
N4276    & SBc         & 33.1$\pm$0.8  &  18.7$\pm$ 4.3 (SDSS/g          )    & 6.4$\pm$0.8   &          ...     &    ...        &  10.2$\pm$0.3  & 5.5$\pm$1.1  (5.3$\pm$1.2)  \\
N4293    & SB(s)0/a    & 30.8$\pm$0.6  &     ...                              &    ...        &  118.0$\pm$17.7  & 6.8$\pm$0.7   &  10.5$\pm$0.2  & 6.3$\pm$1.0  (6.2$\pm$1.0)  \\
N4294    & SB(s)cd     & 31.1$\pm$0.4  &  35.5$\pm$ 2.6 (CFHT/z          )    & 3.5$\pm$0.6   &          ...     &    ...        &   9.7$\pm$0.2  & 4.2$\pm$0.8  (3.8$\pm$0.9)  \\
N4298    & SA(rs)c     & 31.0$\pm$0.3  &  23.5$\pm$ 4.1 (Spitzer/IRAC2   )    & 5.6$\pm$0.8   &   42.0$\pm$ 6.3  & 4.2$\pm$0.8   &  10.2$\pm$0.2  & 5.7$\pm$0.8  (5.4$\pm$0.9)  \\
N4299    & SAB(s)dm    & 31.7$\pm$0.3  &  24.6$\pm$ 7.8 (CFHT/u          )    & 5.4$\pm$1.4   &          ...     &    ...        &   9.5$\pm$0.1  & 3.9$\pm$0.8  (3.3$\pm$0.8)  \\
N4302    & Sc          & 31.4$\pm$0.7  &     ...                              &    ...        &          ...     &    ...        &  10.5$\pm$0.3  & 6.5$\pm$1.1  (6.4$\pm$1.2)  \\
N4303$^a$ & SAB(rs)bc   & 30.6$\pm$0.9  &  14.7$\pm$ 0.9 (GALEX/NUV       )    & 6.6$\pm$0.2   &   96.0$\pm$14.4  & 6.3$\pm$0.7   &  10.6$\pm$0.4  & 6.7$\pm$1.3  (6.6$\pm$1.4)  \\
N4307    & Sb          & 31.7$\pm$0.3  &     ...                              &    ...        &          ...     &    ...        &  10.3$\pm$0.1  & 6.0$\pm$0.8  (5.7$\pm$0.9)  \\
N4312    & SA(rs)ab    & 30.2$\pm$0.2  &     ...                              &    ...        &   69.0$\pm$10.4  & 5.4$\pm$0.7   &   9.6$\pm$0.1  & 4.1$\pm$0.7  (3.6$\pm$0.8)  \\
N4313    & SA(rs)ab    & 30.8$\pm$0.4  &     ...                              &    ...        &   63.0$\pm$ 9.5  & 5.2$\pm$0.8   &  10.0$\pm$0.2  & 5.2$\pm$0.8  (4.9$\pm$0.9)  \\
N4316    & Scd?        & 32.2$\pm$0.3  &     ...                              &    ...        &          ...     &    ...        &  10.3$\pm$0.1  & 5.9$\pm$0.8  (5.6$\pm$0.9)  \\
N4321    & AB(s)bc     & 31.0$\pm$0.4  &  13.4$\pm$ 3.6 (SDSS/g          )    & 7.3$\pm$0.7   &   91.0$\pm$13.7  & 6.1$\pm$0.7   &  10.9$\pm$0.2  & 7.6$\pm$1.2  (7.7$\pm$1.1)  \\
N4330    & Scd         & 31.4$\pm$0.1  &     ...                              &    ...        &          ...     &    ...        &   9.9$\pm$0.1  & 4.8$\pm$0.7  (4.4$\pm$0.8)  \\
N4343    & SA(rs)b     & 32.1$\pm$0.3  &     ...                              &    ...        &          ...     &    ...        &  10.4$\pm$0.1  & 6.1$\pm$0.8  (6.0$\pm$0.9)  \\
N4356    & Scd         & 31.7$\pm$0.6  &     ...                              &    ...        &          ...     &    ...        &  10.0$\pm$0.3  & 5.1$\pm$1.0  (4.7$\pm$1.1)  \\
N4380    & SA(rs)b     & 31.4$\pm$0.4  &  20.0$\pm$ 7.0 (Spitzer/IRAC4   )    & 6.2$\pm$1.2   &   85.0$\pm$12.8  & 6.0$\pm$0.7   &  10.3$\pm$0.2  & 5.8$\pm$0.8  (5.6$\pm$0.9)  \\
N4388$^a$ & SA(s)b      & 31.4$\pm$0.5  &  18.6$\pm$ 2.6 (KPNO 2.3m/K$_s$ )    & 6.9$\pm$0.1   &   99.0$\pm$14.9  & 6.3$\pm$0.7   &  10.5$\pm$0.2  & 6.4$\pm$0.9  (6.2$\pm$1.0)  \\
N4390    & SAB(s)c     & 31.6$\pm$1.0  &  21.7$\pm$ 9.9 (CFHT/u          )    & 5.9$\pm$1.7   &          ...     &    ...        &   9.5$\pm$0.4  & 3.9$\pm$1.3  (3.4$\pm$1.5)  \\
N4394    & SB(r)b      & 31.2$\pm$0.3  &  10.9$\pm$ 6.7 (SDSS/g          )    & 7.7$\pm$1.2   &  120.0$\pm$18.0  & 6.8$\pm$0.7   &  10.4$\pm$0.2  & 6.2$\pm$0.8  (6.0$\pm$0.9)  \\
N4396    & SAd         & 30.7$\pm$0.3  &     ...                              &    ...        &          ...     &    ...        &   9.3$\pm$0.1  & 3.2$\pm$0.8  (2.6$\pm$0.8)  \\
N4402    & Sb          & 30.8$\pm$0.5  &     ...                              &    ...        &          ...     &    ...        &  10.1$\pm$0.2  & 5.3$\pm$0.9  (5.0$\pm$1.0)  \\
N4405    & SA(rs)0/a   & 31.2$\pm$0.4  &     ...                              &    ...        &          ...     &    ...        &   9.9$\pm$0.2  & 4.8$\pm$0.8  (4.4$\pm$0.9)  \\
N4411b   & SAB(s)cd    & 31.7$\pm$0.8  &  23.4$\pm$ 1.6 (CFHT/u          )    & 5.6$\pm$0.4   &          ...     &    ...        &   9.7$\pm$0.3  & 4.5$\pm$1.1  (4.0$\pm$1.2)  \\
N4412    & SB(r)b      & 32.8$\pm$0.8  &  10.8$\pm$ 3.9 (SDSS/u          )    & 7.7$\pm$0.8   &          ...     &    ...        &  10.3$\pm$0.3  & 6.0$\pm$1.1  (5.8$\pm$1.2)  \\
N4413    & SB(rs)ab    & 31.0$\pm$0.2  &  29.8$\pm$ 2.7 (GALEX/FUV \& NUV)    & 4.5$\pm$0.6   &          ...     &    ...        &   9.6$\pm$0.1  & 4.1$\pm$0.7  (3.6$\pm$0.8)  \\
N4416    & SB(rs)cd    & 33.1$\pm$0.8  &  32.0$\pm$ 9.7 (SDSS/u          )    & 4.1$\pm$1.7   &          ...     &    ...        &  10.3$\pm$0.3  & 5.9$\pm$1.1  (5.7$\pm$1.2)  \\
N4419    & SB(s)a      & 31.0$\pm$0.5  &     ...                              &    ...        &  102.0$\pm$15.3  & 6.4$\pm$0.7   &  10.4$\pm$0.2  & 6.3$\pm$0.9  (6.1$\pm$1.0)  \\
N4424    & SB(s)a      & 30.6$\pm$1.0  &  16.9$\pm$ 2.4 (SDSS/r          )    & 6.7$\pm$0.5   &   57.0$\pm$ 8.6  & 5.0$\pm$0.8   &   9.8$\pm$0.4  & 4.6$\pm$1.3  (4.2$\pm$1.5)  \\
N4429    & SA0(r)      & 31.0$\pm$0.6  &     ...                              &    ...        &  173.0$\pm$26.0  & 7.7$\pm$0.6   &  10.8$\pm$0.2  & 7.2$\pm$1.1  (7.2$\pm$1.1)  \\
N4430    & SB(rs)b     & 31.2$\pm$0.9  &  20.1$\pm$ 8.2 (SDSS/g          )    & 6.1$\pm$1.4   &          ...     &    ...        &   9.9$\pm$0.4  & 4.8$\pm$1.2  (4.4$\pm$1.4)  \\
N4438    & SA(s)0/a    & 30.2$\pm$0.7  &     ...                              &    ...        &  135.0$\pm$20.2  & 7.1$\pm$0.7   &  10.4$\pm$0.3  & 6.3$\pm$1.0  (6.1$\pm$1.1)  \\
N4445    & Sab         & 31.2$\pm$0.5  &     ...                              &    ...        &          ...     &    ...        &   9.7$\pm$0.2  & 4.3$\pm$0.8  (3.8$\pm$0.9)  \\
N4450    & SA(s)ab     & 30.8$\pm$0.5  &  10.1$\pm$ 2.4 (Spitzer/IRAC4   )    & 7.8$\pm$0.5   &  132.0$\pm$19.8  & 7.0$\pm$0.7   &  10.7$\pm$0.2  & 6.9$\pm$1.0  (6.9$\pm$1.0)  \\
N4451    & Sbc         & 32.0$\pm$0.4  &     ...                              &    ...        &          ...     &    ...        &   9.9$\pm$0.2  & 5.0$\pm$0.8  (4.6$\pm$0.9)  \\
N4457    & SAB(s)0/a   & 30.3$\pm$0.5  &     ...                              &    ...        &  113.0$\pm$17.0  & 6.7$\pm$0.7   &  10.2$\pm$0.2  & 5.6$\pm$0.9  (5.3$\pm$1.0)  \\
N4469    & SB(s)0/a    & 31.1$\pm$0.8  &     ...                              &    ...        &  106.0$\pm$15.9  & 6.5$\pm$0.7   &  10.4$\pm$0.3  & 6.1$\pm$1.1  (5.9$\pm$1.2)  \\
N4470    & Sa          & 31.2$\pm$0.7  &  29.2$\pm$14.9 (SDSS/gri        )    & 4.6$\pm$2.6   &   90.0$\pm$13.5  & 6.1$\pm$0.7   &   9.6$\pm$0.3  & 4.1$\pm$1.0  (3.6$\pm$1.1)  \\
N4480    & SAB(s)c     & 33.0$\pm$0.2  &  18.4$\pm$ 0.9 (Spitzer/IRAC2   )    & 6.4$\pm$0.3   &          ...     &    ...        &  10.4$\pm$0.1  & 6.3$\pm$0.8  (6.1$\pm$0.8)  \\
N4498    & SAB(s)d     & 31.0$\pm$0.5  &  22.2$\pm$10.5 (GALEX/NUV       )    & 5.8$\pm$1.8   &          ...     &    ...        &   9.7$\pm$0.2  & 4.3$\pm$0.8  (3.8$\pm$0.9)  \\
N4501$^a$ & SA(rs)b     & 31.2$\pm$0.5  &  12.2$\pm$ 3.4 (GALEX/NUV       )    & 7.1$\pm$0.8   &  166.0$\pm$24.9  & 7.6$\pm$0.6   &  11.1$\pm$0.2  & 8.0$\pm$1.6  (8.2$\pm$1.4)  \\
N4519    & SB(rs)d     & 31.6$\pm$0.8  &  23.9$\pm$13.4 (GALEX/FUV       )    & 5.5$\pm$2.3   &          ...     &    ...        &   9.9$\pm$0.3  & 4.7$\pm$1.1  (4.4$\pm$1.2)  \\
N4522    & SB(s)cd     & 31.1$\pm$0.5  &     ...                              &    ...        &          ...     &    ...        &   9.7$\pm$0.2  & 4.3$\pm$0.9  (3.8$\pm$1.0)  \\
N4527    & SAB(s)bc    & 30.7$\pm$0.4  &  12.8$\pm$ 3.1 (Spitzer/IRAC2   )    & 7.4$\pm$0.6   &  135.0$\pm$20.2  & 7.1$\pm$0.7   &  10.6$\pm$0.2  & 6.8$\pm$0.9  (6.7$\pm$1.0)  \\
N4532    & IBm         & 30.6$\pm$0.5  &     ...                              &    ...        &          ...     &    ...        &   9.6$\pm$0.2  & 4.0$\pm$0.9  (3.5$\pm$1.0)  \\
N4535    & SAB(s)c     & 30.9$\pm$0.5  &  21.9$\pm$ 4.0 (Spitzer/IRAC2   )    & 5.8$\pm$0.8   &  102.0$\pm$15.3  & 6.4$\pm$0.7   &  10.6$\pm$0.2  & 6.7$\pm$1.0  (6.6$\pm$1.0)  \\
N4536    & SAB(rs)bc   & 30.9$\pm$0.4  &  20.2$\pm$ 2.3 (Spitzer/IRAC2   )    & 6.1$\pm$0.5   &  111.0$\pm$16.7  & 6.6$\pm$0.7   &  10.4$\pm$0.2  & 6.3$\pm$0.9  (6.1$\pm$0.9)  \\
N4548    & SB(rs)b     & 31.0$\pm$0.3  &  14.6$\pm$ 3.6 (SDSS/u          )    & 7.1$\pm$0.7   &  122.0$\pm$18.3  & 6.8$\pm$0.7   &  10.7$\pm$0.2  & 6.9$\pm$0.9  (6.8$\pm$0.9)  \\
N4567    & SA(rs)bc    & 31.8$\pm$0.5  &  13.4$\pm$ 1.4 (SDSS/g          )    & 7.3$\pm$0.4   &   66.0$\pm$ 9.9  & 5.3$\pm$0.7   &  10.5$\pm$0.2  & 6.4$\pm$0.9  (6.3$\pm$1.0)  \\
N4568    & SA(rs)bc    & 31.5$\pm$0.6  &  21.7$\pm$ 5.2 (Spitzer/IRAC1   )    & 5.9$\pm$0.9   &   88.0$\pm$13.2  & 6.0$\pm$0.7   &  10.7$\pm$0.2  & 7.0$\pm$1.1  (7.0$\pm$1.1)  \\
N4569    & SAB(rs)ab   & 30.4$\pm$0.6  &  14.3$\pm$ 6.8 (SDSS/g          )    & 7.1$\pm$1.2   &  139.0$\pm$20.9  & 7.2$\pm$0.7   &  10.6$\pm$0.2  & 6.8$\pm$1.0  (6.7$\pm$1.1)  \\
N4571    & SA(r)d      & 31.0$\pm$0.3  &  11.2$\pm$ 7.4 (HST/F555W       )    & 7.7$\pm$1.3   &          ...     &    ...        &  10.1$\pm$0.1  & 5.5$\pm$0.8  (5.2$\pm$0.8)  \\
	\hline
	\end{tabular}}
\end{table*}

\begin{table*}
\label{Tab_App2}
\centering
\caption{Continued.} 
\resizebox{\textwidth}{!}{%
\begin{tabular}{llccccccc} 
\hline
 Galaxy  &  Type       &   Dist.Mod.   & $|\phi|$ (band) &  $\log M_{\rm bh}(|\phi|)$  &   $\sigma$      &  $\log M_{\rm bh}(\sigma)$  & $\log M_{\rm *,gal}$ & $\log M_{\rm bh}(M_{\rm *,gal})$ \\
         &             &               & [deg]           &     [dex]     &  km s$^{-1}$      &   [dex]       &   [dex]        &    [dex]   \\      
	\hline
N4579    & SAB(rs)b    & 31.3$\pm$0.4  &   6.1$\pm$ 3.9 (Spitzer/IRAC4   )    & 8.5$\pm$0.8   &  166.0$\pm$24.9  & 7.6$\pm$0.6   &  11.1$\pm$0.2  & 7.9$\pm$1.4  (8.0$\pm$1.3)  \\
N4580    & SAB(rs)a    & 31.4$\pm$1.1  &  20.2$\pm$ 6.2 (SDSS/griz       )    & 6.1$\pm$1.1   &          ...     &    ...        &  10.1$\pm$0.4  & 5.4$\pm$1.3  (5.1$\pm$1.5)  \\
N4606    & SB(s)a      & 31.0$\pm$0.5  &     ...                              &    ...        &          ...     &    ...        &   9.8$\pm$0.2  & 4.5$\pm$0.9  (4.0$\pm$1.0)  \\
N4607    & SBb?        & 31.0$\pm$0.9  &     ...                              &    ...        &          ...     &    ...        &   9.7$\pm$0.4  & 4.4$\pm$1.2  (4.0$\pm$1.3)  \\
N4639    & SAB(rs)bc   & 31.8$\pm$0.2  &  17.5$\pm$ 4.9 (SDSS/g          )    & 6.6$\pm$0.9   &   91.0$\pm$13.7  & 6.1$\pm$0.7   &  10.3$\pm$0.1  & 6.0$\pm$0.8  (5.8$\pm$0.8)  \\
N4647    & SAB(rs)c    & 31.2$\pm$0.3  &  21.4$\pm$ 2.6 (Spitzer/IRAC2   )    & 5.9$\pm$0.5   &   49.0$\pm$ 7.4  & 4.6$\pm$0.8   &  10.2$\pm$0.2  & 5.6$\pm$0.8  (5.3$\pm$0.9)  \\
N4651    & SA(rs)c     & 31.7$\pm$0.8  &  14.3$\pm$ 1.0 (CFHT/u          )    & 7.1$\pm$0.4   &  101.0$\pm$15.2  & 6.4$\pm$0.7   &  10.6$\pm$0.3  & 6.7$\pm$1.1  (6.6$\pm$1.3)  \\
N4654    & SAB(rs)cd   & 30.8$\pm$0.5  &  25.5$\pm$ 9.3 (Spitzer/IRAC3   )    & 5.2$\pm$1.6   &   48.0$\pm$ 7.2  & 4.6$\pm$0.8   &  10.4$\pm$0.2  & 6.1$\pm$0.9  (5.9$\pm$1.0)  \\
N4689    & SA(rs)bc    & 31.0$\pm$0.4  &  24.1$\pm$ 7.2 (Spitzer/IRAC2   )    & 5.4$\pm$1.3   &   41.0$\pm$ 6.2  & 4.2$\pm$0.8   &  10.1$\pm$0.2  & 5.5$\pm$0.8  (5.2$\pm$0.9)  \\
N4698    & SA(s)ab     & 31.6$\pm$0.8  &     ...                              &    ...        &  137.0$\pm$20.6  & 7.1$\pm$0.7   &  10.8$\pm$0.3  & 7.1$\pm$1.2  (7.1$\pm$1.3)  \\
N4713    & SAB(rs)d    & 30.6$\pm$0.8  &  35.4$\pm$10.8 (HST F606W       )    & 3.5$\pm$1.9   &   23.0$\pm$ 3.5  & 2.8$\pm$1.0   &   9.6$\pm$0.3  & 4.0$\pm$1.1  (3.5$\pm$1.2)  \\
\hline
\end{tabular}}

$^a$ Directly measured black hole masses exist for these three galaxies (see Section~\ref{SecData}).
The galaxy and black hole masses in this table are in units of solar masses.  
The three sets of predicted black hole mass are based on: the spiral arm's
pitch angle $\phi$ using equation~\ref{eq1b}; 
the galaxy's central velocity dispersion $\sigma$ using equation~\ref{eq2}; 
and the galaxy's stellar mass $M_{\rm *,galaxy}$ derived using
equation~\ref{eq3} (and equation~\ref{eq3B}).

\end{table*}


\bsp	
\label{lastpage}

\begin{thebibliography}{199}

 \bibitem[Abbott et al.(2016)]{2016PhRvL.116f1102A} Abbott, B.~P., Abbott, R.,
  Abbott, T.~D., et al.\ 2016, Physical Review Letters, 116, 061102 
 \bibitem[Abbott et al.(2017)]{2017PhRvL.118v1101A} Abbott, B.~P., Abbott, R.,
  Abbott, T.~D., et al.\ 2017, Physical Review Letters, 118, 221101 
 \bibitem[Alam et al.(2015)]{2015ApJS..219...12A} Alam, S., Albareti, F.~D.,
  Allende Prieto, C., et al.\ 2015, \apjs, 219, 12
 \bibitem[Alexander \& Natarajan(2014)]{2014Sci...345.1330A} Alexander, T., \&
   Natarajan, P.\ 2014, Science, 345, 1330 
 \bibitem[Alvarez et al.(2009)]{2009ApJ...701L.133A} Alvarez, M.~A., Wise,
  J.~H., \& Abel, T.\ 2009, \apjl, 701, L133 
 \bibitem[Anderson \& van der Marel(2010)]{anderson10} Anderson J., van der Marel R.P.\  2010, ApJ, 710, 1032
 \bibitem[Anza et al.(2005)]{Anza:2005} Anza, S., Armano, M., Balaguer, E., et
  al.\ 2005, Classical and Quantum Gravity, 22, 125
 \bibitem[Argyres et al.(1998)]{1998PhLB..441...96A} Argyres, P.~C.,
  Dimopoulos, S., \& March-Russell, J.\ 1998, Physics Letters B, 441, 96 
 \bibitem[Arnaud(1996)]{arnaud96} Arnaud K.A., 1996, ASPC, 101, 17
 \bibitem[Asada et al.(2017)]{2017arXiv170504776A} Asada, K., Kino, M., Honma,
   M., et al.\ 2017, arXiv:1705.04776 
 \bibitem[Aso et al.(2013)]{2013PhRvD..88d3007A} Aso, Y., Michimura, Y.,
  Somiya, K., et al.\ 2013, Physical Review D, 88, 043007 
 \bibitem[Bahcall et al.(1972)]{BKS72}Bahcall J.N., Kozlovsky B.Z., Salpeter E.E., 1972, ApJ, 171, 467
 \bibitem[Baldassare et al.(2015)]{2015ApJ...809L..14B} Baldassare, V.~F.,
   Reines, A.~E., Gallo, E., \& Greene, J.~E.\ 2015, \apjl, 809, L14
 \bibitem[Baldwin et al.(1981)]{1981PASP...93....5B} Baldwin, J.~A., Phillips,
   M.~M., \& Terlevich, R.\ 1981, \pasp, 93, 5 
 \bibitem[Bean \& Magueijo(2002)]{2002PhRvD..66f3505B} Bean, R., \& Magueijo,
   J.\ 2002, \prd, 66, 063505 
 \bibitem[Bekenstein(1973)]{1973ApJ...183..657B} Bekenstein, J.~D.\ 1973, \apj,
  183, 657 
 \bibitem[Belczynski et al.(2010)]{2010ApJ...714.1217B} Belczynski, K., Bulik,
  T., Fryer, C.~L., et al.\ 2010, \apj, 714, 1217 
 \bibitem[Bell et al.(2003)]{2003ApJS..149..289B} Bell, E.~F., McIntosh,
  D.~H., Katz, N., \& Weinberg, M.~D.\ 2003, \apjs, 149, 289 
 \bibitem[Bell \& de Jong(2001)]{2001ApJ...550..212B} Bell, E.~F., \& de Jong,
   R.~S.\ 2001, \apj, 550, 212 
 \bibitem[Berrier et al.(2013)]{2013ApJ...769..132B} Berrier, J.~C., Davis,
   B.~L., Kennefick, D., et al.\ 2013, \apj, 769, 132 
 \bibitem[Blandford \& McKee(1982)]{BaM82}Blandford R.D., McKee C.F., 1982, ApJ, 255, 419
 \bibitem[Bond et al.(1984)]{1984ApJ...280..825B} Bond, J.~R., Arnett, W.~D.,
  \& Carr, B.~J.\ 1984, \apj, 280, 825 
 \bibitem[Bromm \& Loeb(2003)]{2003AIPC..666...73B} Bromm, V., \& Loeb,
   A.\ 2003, The Emergence of Cosmic Structure, 666, 73 
 \bibitem[Carr et al.(1984)]{1984ApJ...277..445C} Carr, B.~J., Bond, J.~R., \&
  Arnett, W.~D.\ 1984, \apj, 277, 445 
 \bibitem[Carr et al.(2010)]{2010PhRvD..81j4019C} Carr, B.~J., Kohri, K.,
  Sendouda, Y., \& Yokoyama, J.\ 2010, \prd, 81, 104019 
 \bibitem[Cash(1979)]{cash79} Cash W.\ 1979, ApJ, 228, 939
 \bibitem[Chabrier(2003)]{2003PASP..115..763C} Chabrier, G.\ 2003, \pasp, 115,
   763 
 \bibitem[Clark et al.(1975)]{1975ApJ...199L..93C} Clark, G.~W., Markert, T.~H., \& Li, F.~K.\ 1975, \apjl, 199, L93 
 \bibitem[Colbert \& Mushotzky(1999)]{1999ApJ...519...89C} Colbert, E.~J.~M.,
   \& Mushotzky, R.~F.\ 1999, \apj, 519, 89
 \bibitem[Conroy et al.(2009)]{2009ApJ...699..486C} Conroy, C., Gunn, J.~E.,
   \& White, M.\ 2009, \apj, 699, 486 
 \bibitem[Cseh et al.(2015)]{2015MNRAS.446.3268C} Cseh, D., Webb, N.~A., Godet,
   O., et al.\ 2015, \mnras, 446, 3268 
 \bibitem[Davies et al.(1983)]{DEF83}Davies, R.L., Efstathiou, G., Fall, S.M., et al.\ 1983, ApJ, 266, 41
 \bibitem[Davis et al.(2012)]{2012ApJS..199...33D} Davis, B.~L., Berrier,
   J.~C., Shields, D.~W., et al.\ 2012, \apjs, 199, 33 
 \bibitem[Davis et al.(2018a)]{2018aDavis} Davis, B.~L., Graham,
  A.~W., \& Cameron, E.\ 2018a, ApJ, submitted, arXiv:1810.04887
 \bibitem[Davis et al.(2018b)]{2018bDavis} Davis, B.~L., Graham,
  A.~W., \& Cameron, E.\ 2018b, ApJ, submitted, arXiv:1810.04888
 \bibitem[Davis et al.(2017)]{2017MNRAS.471.2187D} Davis, B.~L., Graham, A.~W.,
  \& Seigar, M.~S.\ 2017, \mnras, 471, 2187 
 \bibitem[Davis et al.(2011)]{davis11} Davis S.W., Narayan R., Zhu Y., Barret
   D., Farrell S.A., Godet O., Servillat M., Webb N.A. 2011, ApJ, 734, 111
 \bibitem[Decarli et al.(2007)]{decarli07} Decarli R., Gavazzi G., Arosio I.,
   Cortese L., Boselli A., Bonfanti C., Colpi M. 2007, MNRAS, 381, 136
 \bibitem[den Brok et al.(2015)]{2015ApJ...809..101D} den Brok, M., Seth,
   A.~C., Barth, A.~J., et al.\ 2015, \apj, 809, 101 
 \bibitem[de Vaucoulers et al.(1991)]{deV91} de Vaucouleurs, G., de
   Vaucouleurs, A., Corwin, H.~G.~Jr, Buta R.~J., Paturel G., \& Fouque
   P.\ 1991, Third Reference Catalogue of Bright Galaxies. Springer-Verlag,
   Berlin (RC3)
 \bibitem[Doroshkevich et al..(1967)]{DZN67} Doroshkevich, A. G., Zel'dovich,
   Ya. B., and Novikov, I. D. 1967, Astr. Zh., 44, 295 (Transl., in Soviet
   Astr., 11, 233, 1967).
 \bibitem[Drinkwater et al.(2003)]{Drinkwater}Drinkwater, M. J., Gregg, M. D.,
  Hilker, M., et al., 2003, Nature, 423, 519
 \bibitem[Driver et al.(2007)]{2007MNRAS.379.1022D} Driver, S.~P., Popescu,
   C.~C., Tuffs, R.~J., et al.\ 2007, \mnras, 379, 1022 
 \bibitem[Driver et al.(2008)]{2008ApJ...678L.101D} Driver, S.~P., Popescu,
    C.~C., Tuffs, R.~J., et al.\ 2008, \apjl, 678, L101 
 \bibitem[Dudik et al.(2005)]{dudik05} Dudik R.P., Satyapal S., Gliozzi M., Sambruna R.M.\ 2005, ApJ, 620, 113	
 \bibitem[Falcke et al.(2004)]{2004A&A...414..895F} Falcke, H., K{\"o}rding,
  E., \& Markoff, S.\ 2004, \aap, 414, 895
 \bibitem[Farrell et al.(2009)]{Far09}Farrell, S.A., Webb, N.A., Barret, D.,
   Godet, O., \& Rodrigues, J.M.\ 2009, Nature, 460, 73
 \bibitem[Farrell et al.(2014)]{farrell14} Farrell S.A.,
   Servillat M., Gladstone J.C., et al.\ 2014, \mnras, 437, 1208
 \bibitem[Favata et al.(2004)]{2004ApJ...607L...5F} Favata, M., Hughes, S.~A.,
  \& Holz, D.~E.\ 2004, \apjl, 607, L5
 \bibitem[Feng \& Soria(2011)]{feng11} Feng H., Soria R.\ 2011, NewAR, 55, 166 
 \bibitem[Ferrarese \& Ford(2005)]{2005SSRv..116..523F} Ferrarese, L., \&
   Ford, H.\ 2005, \ssr, 116, 523 
 \bibitem[Filippenko \& Ho(2003)]{2003ApJ...588L..13F} Filippenko, A.~V., \&
   Ho, L.~C.\ 2003, \apjl, 588, L13 
 \bibitem[Fraser et al.(2017)]{2017MNRAS.468..418F} Fraser, M., Casey, A.~R.,
  Gilmore, G., Heger, A., \& Chan, C.\ 2017, \mnras, 468, 418 
 \bibitem[Fruscione et al.(2006)]{fruscione06} Fruscione A., et al.\ 2006, SPIE, 6270, 1
\bibitem[Gaia Collaboration et al.(2018)]{2018A&A...616A...1G} Gaia
  Collaboration, Brown, A.~G.~A., Vallenari, A., et al.\ 2018, \aap, 616, A1
 \bibitem[Gallo et al.(2008)]{2008ApJ...680..154G} Gallo, E., Treu, T., Jacob,
   J., et al.\ 2008, \apj, 680, 154-168 
 \bibitem[Gallo et al.(2010)]{2010ApJ...714...25G} Gallo, E., Treu, T.,
   Marshall, P.~J., et al.\ 2010, \apj, 714, 25 
 \bibitem[Gavazzi et al.(2003)]{2003A&A...400..451G} Gavazzi, G., Boselli, A.,
   Donati, A., Franzetti, P., \& Scodeggio, M.\ 2003, \aap, 400, 451
 \bibitem[Godet et al.(2012)]{godet12} Godet O., et al.\ 2012, ApJ, 752, 34
 \bibitem[Gonz{\'a}lez Delgado et al.(2008)]{2008AJ....135..747G} Gonz{\'a}lez
  Delgado, R.~M., P{\'e}rez, E., Cid Fernandes, R., \& Schmitt, H.\ 2008, \aj,
  135, 747 
 \bibitem[Graham(2016a)]{2016ASSL..418..263G} Graham, A.~W.\ 2016a, in Galactic
  Bulges, E. Laurikainen, R.F. Peletier, and D.A. Gadotti (eds.), Springer
  Publishing, Astrophysics and Space Science Library, v.418, p.263-313 
 \bibitem[Graham(2016b)]{2016IAUS..312..269G} Graham, A.~W.\ 2016b, Star
   Clusters and Black Holes in Galaxies across Cosmic Time, 312, 269
 \bibitem[Graham et al.(2016)]{2016ApJ...818..172G} Graham, A.~W., Ciambur,
  B.~C., \& Soria, R.\ 2016, \apj, 818, 172 
 \bibitem[Graham \& Driver(2007)]{2007ApJ...655...77G} Graham, A.~W., \&
   Driver, S.~P.\ 2007, \apj, 655, 77 
 \bibitem[Graham et al.(2003)]{2003AJ....126.1787G} Graham, A.~W., Jerjen, H.,
   \& Guzm{\'a}n, R.\ 2003, \aj, 126, 1787 
 \bibitem[Graham et al.(2017)]{2017ApJ...840...68G} Graham, A.~W., Janz, J.,
   Penny, S.~J., et al.\ 2017, \apj, 840, 68 
 \bibitem[Graham \& Scott(2013)]{2013ApJ...764..151G} Graham, A.~W., \& Scott,
   N.\ 2013, \apj, 764, 151 
 \bibitem[Graham \& Scott(2015)]{2015ApJ...798...54G} Graham, A.~W., \& Scott,
  N.\ 2015, \apj, 798, 54 
 \bibitem[Graham \& Soria(2018)]{GaS18} Graham, A.~W., \& Soria, R.\ 2018,
   MNRAS, submitted (Paper I)
 \bibitem[Graham \& Spitler(2009)]{2009MNRAS.397.2148G} Graham, A.~W., \&
  Spitler, L.~R.\ 2009, \mnras, 397, 2148 
 \bibitem[Greenhill et al.(2003)]{2003ApJ...590..162G} Greenhill, L.~J.,
   Booth, R.~S., Ellingsen, S.~P., et al.\ 2003, ApJ, 590, 162
 \bibitem[Grobov et al.(2011)]{2011GrCo...17..181G} Grobov, A.~V., Rubin,
   S.~G., Samarchenko, D.~A., \& Zhizhin, E.~D.\ 2011, Gravitation and
   Cosmology, 17, 181 
 \bibitem[G\"urkan et al.(2004)]{gurkan04} G\"urkan M.A., Freitag M., Rasio F.A., 2004, ApJ, 604, 632
 \bibitem[Haehnelt \& Rees(1993)]{1993MNRAS.263..168H} Haehnelt, M.~G., \&
  Rees, M.~J.\ 1993, \mnras, 263, 168 
 \bibitem[Heckman \& Best(2014)]{2014ARA&A..52..589H} Heckman, T.~M., \& Best,
  P.~N.\ 2014, ARA\&A, 52, 589
 \bibitem[Herrmann et al.(2007)]{2007ApJ...661..430H} Herrmann, F., Hinder, I.,
  Shoemaker, D., Laguna, P., \& Matzner, R.~A.\ 2007, \apj, 661, 430 
 \bibitem[Ho et al.(1997)]{ho97} Ho L.C., Filippenko A.V., Sargent W.L.\ 1997, ApJS, 112, 315
 \bibitem[Hobbs et al.(2010)]{2010CQGra..27h4013H} Hobbs, G., Archibald, A.,
  Arzoumanian, Z., et al.\ 2010, Classical and Quantum Gravity, 27, 084013
 \bibitem[Hubble(1926)]{Hub26}Hubble, E.\ 1926, ApJ, 64, 321
   \bibitem[Hubble(1936)]{Hub36}Hubble, E.P.\ 1936a, Realm of the Nebulae, by
     E.P.\ Hubble, New Haven: Yale University Press
 \bibitem[Humphreys et al.(2016)]{2016A&A...592L..13H} Humphreys, E.~M.~L.,
   Vlemmings, W.~H.~T., Impellizzeri, C.~M.~V., et al.\ 2016, \aap, 592, L13
 \bibitem[Ishihara et al.(2010)]{Ishihara2010} Ishihara, D., Onaka, T., Kataza, H., et al.\ 2010, \aap, 514, A1 
 \bibitem[Iwasawa et al.(2000)]{2000MNRAS.318..879I} Iwasawa, K., Fabian,
   A.~C., Almaini, O., et al.\ 2000, \mnras, 318, 879 
 \bibitem[Iye et al.(1982)]{1982ApJ...256..103I} Iye, M., Okamura, S., Hamabe,
   M., \& Watanabe, M.\ 1982, \apj, 256, 103 
 \bibitem[Jarrett et al.(2000)]{2MASS}Jarrett, T.H., Chester, T., Cutri, R., et al.\ 2000, AJ, 119, 2498 (2MASS)
\bibitem[Jeans(1919)]{Jeans1919}Jeans J. 1919. Problems of Cosmogony and
  Stellar Dynamics, Cambridge: Cambridge Univ. Press
  \bibitem[Jeans(1928)]{Jeans1928}Jeans, J.H.\ 1928, Astronomy \&Cosmogony,
    (Cambridge: Cambridge University Press), p.332
 \bibitem[Jiang et al.(2018)]{Jiang2018}Jiang, N., Wang, T., Zhou, H., et
   al.\ 2018, ApJ, in press 
 \bibitem[Johnson \& Bromm(2007)]{2007MNRAS.374.1557J} Johnson, J.~L., \&
   Bromm, V.\ 2007, \mnras, 374, 1557 
 \bibitem[Kaaret \& Feng(2013)]{2013ApJ...770...20K} Kaaret, P., \& Feng,
  H.\ 2013, \apj, 770, 20
 \bibitem[Kaaret et al.(2017)]{kaaret17} Kaaret P., Feng, H., Roberts, T.P., 2017, ARA\&A, 55, 303
 \bibitem[Kalnajs(1975)]{1975dgs..conf..103K} Kalnajs, A.~J.\ 1975, La
   Dynamique des galaxies spirales, 241, 103 
 \bibitem[Kawada et al.(2007)]{Kawada2007} Kawada, M., Baba, H., Barthel, P.~D., et al.\ 2007, \pasj, 59, S389
 \bibitem[Kawamura et al.(2011)]{2011CQGra..28i4011K} Kawamura, S., Ando, M.,
  Seto, N., et al.\ 2011, Classical and Quantum Gravity, 28, 094011 
 \bibitem[Kennicutt(1981)]{1981AJ.....86.1847K} Kennicutt, R.~C., Jr.\ 1981,
   \aj, 86, 1847 
 \bibitem[Kewley et al.(2001)]{2001ApJ...556..121K} Kewley, L.~J., Dopita,                                       
   M.~A., Sutherland, R.~S., Heisler, C.~A., \& Trevena, J.\ 2001, \apj, 556, 121
 \bibitem[Kirby et al.(2008)]{2008AJ....136.1866K} Kirby, E.~M., Jerjen, H.,
   Ryder, S.~D., \& Driver, S.~P.\ 2008, \aj, 136, 1866 
 \bibitem[Kiziltan et al.(2017)]{kiziltan2017} Kiziltan B., Baumgardt H., Loeb A.\ 2017, Natur, 542, 203
 \bibitem[Koliopanos et al.(2017)]{2017A&A...601A..20K} Koliopanos, F.,
  Ciambur, B.~C., Graham, A.~W., et al.\ 2017, \aap, 601, A20 
 \bibitem[Komossa(2013)]{Komossa:2013} Komossa, S.\ 2013, IAU Symposium, 290,
   53
 \bibitem[Komossa et~al.(2009)]{Komossa:2009} {Komossa}, S., {Zhou}, H.,
   {Rau}, A., {et~al.} 2009, ApJ, 701, 105
 \bibitem[Kormendy \& Ho(2013)]{2013ARA&A..51..511K} Kormendy, J., \& Ho,
   L.~C.\ 2013, \araa, 51, 511 
 \bibitem[Koushiappas et al.(2004)]{2004MNRAS.354..292K} Koushiappas, S.~M.,
  Bullock, J.~S., \& Dekel, A.\ 2004, \mnras, 354, 292 
 \bibitem[Kraft et al.(1991)]{kraft91} Kraft R.P., Burrows D.N., Nousek J.A. 1991, \apj, 374, 344
 \bibitem[Krakow et al.(1982)]{1982AJ.....87..203K} Krakow, W., Huntley,
   J.~M., \& Seiden, P.~E.\ 1982, \aj, 87, 203 
 \bibitem[Kramer \& Champion(2013)]{2013CQGra..30v4009K} Kramer, M., \&
  Champion, D.~J.\ 2013, Classical and Quantum Gravity, 30, 224009
 \bibitem[Laine et al.(2014)]{2014MNRAS.444.3015L} Laine, S., Knapen, J.~H.,
  Mu{\~n}oz-Mateos, J.-C., et al.\ 2014, \mnras, 444, 3015 
 \bibitem[Larson(1970)]{1970MNRAS.150...93L} Larson, R.~B.\ 1970, \mnras, 150,
  93 
 \bibitem[Larson(1998)]{1998MNRAS.301..569L} Larson, R.~B.\ 1998, \mnras, 301,
  569 
 \bibitem[Lin et al.(2018)]{2018NatAs.tmp...73L} Lin, D., Strader, J.,
   Carrasco, E.~R., et al.\ 2018, Nature Astronomy, 2, 656 
 \bibitem[Liu et al.(2012)]{2012ApJ...745...89L} Liu, J., Orosz, J., \&
  Bregman, J.~N.\ 2012, \apj, 745, 89
 \bibitem[Loeb \& Rasio(1994)]{1994ApJ...432...52L} Loeb, A., \& Rasio,
  F.~A.\ 1994, \apj, 432, 52 
 \bibitem[Madau \& Rees(2001)]{2001ApJ...551L..27M} Madau, P., \& Rees,
  M.~J.\ 2001, \apjl, 551, L27 
 \bibitem[Mapelli et al.(2012)]{2012A&A...542A.102M} Mapelli, M., Ripamonti,
  E., Vecchio, A., Graham, A.~W., \& Gualandris, A.\ 2012, \aap, 542, A102 
 \bibitem[Mapelli et al.(2013)]{mapelli13} Mapelli M., Annibali F., Zampieri L., Soria R.\ 2013, A\&A, 559, 124
 \bibitem[Matkovi\'c \& Guzm\'an(2005)]{MaG05}Matkovi\'c, A., \& Guzm\'an, R.\ 2005, MNRAS, 362, 289
 \bibitem[Mayer et al.(2010)]{2010Natur.466.1082M} Mayer, L., Kazantzidis, S.,
   Escala, A., \& Callegari, S.\ 2010, \nat, 466, 1082  
 \bibitem[McNamara(2013)]{McNamara:2013} McNamara, P.~W.\ 2013, International
   Journal of Modern Physics D, 22, 41001
 \bibitem[Meidt et al.(2014)]{2014ApJ...788..144M} Meidt, S.~E., Schinnerer,
   E., van de Ven, G., et al.\ 2014, \apj, 788, 144 
 \bibitem[Merloni et al.(2003)]{merloni03} Merloni A., Heinz S., di Matteo T.\ 2003, MNRAS, 345, 1057 
 \bibitem[Merritt et al.(2009)]{Merritt}Merritt, D., Schnittman, J. D., \&
  Komossa, S. 2009, ApJ, 699, 1690
 \bibitem[Mezcua(2017)]{2017IJMPD..2630021M} Mezcua, M.\ 2017, International
   Journal of Modern Physics D, 26, 1730021 
 \bibitem[Mezcua et al.(2015)]{2015MNRAS.448.1893M} Mezcua, M., Roberts,
  T.~P., Lobanov, A.~P., \& Sutton, A.~D.\ 2015, \mnras, 448, 1893
 \bibitem[Mezcua et al.(2016)]{2016ApJ...817...20M} Mezcua, M., Civano, F.,
   Fabbiano, G., Miyaji, T., \& Marchesi, S.\ 2016, \apj, 817, 20 
 \bibitem[Mezcua et al.(2018)]{2018MNRAS.478.2576M} Mezcua, M., Civano, F.,
   Marchesi, S., et al.\ 2018, \mnras, 478, 2576 
 \bibitem[Miller \& Colbert(2004)]{2004IJMPD..13....1M} Miller, M.~C., \&
  Colbert, E.~J.~M.\ 2004, International Journal of Modern Physics D, 13, 1 
 \bibitem[Miller et al.(2003)] {miller03} Miller J.M., Fabbiano G., Miller M.C., Fabian A.C. 2003, ApJ, 585, L37
 \bibitem[Miller et al.(2013)]{2013:Miller} Miller, J. M., Walton, D. J., King,
  A. L., et al., 2013, ApJL, 776, L36
 \bibitem[Miller-Jones et al.(2012)]{miller-jones12} Miller-Jones J.C.A., et al.\ 2012, ApJ, 755, L1
 \bibitem[Miyoshi et al.(1995)]{1995Natur.373..127M} Miyoshi, M., Moran, J.,
  Herrnstein, J., et al.\ 1995, Nature, 373, 127 
 \bibitem[Nagar et al.(2002)]{2002A&A...392...53N} Nagar, N.~M., Falcke, H.,
   Wilson, A.~S., \& Ulvestad, J.~S.\ 2002, \aap, 392, 53 
 \bibitem[Nagar(2013)]{2013PhRvD..88l1501N} Nagar, A.\ 2013, \prd, 88, 121501 
 \bibitem[Narayan \& Yi(1995)]{1995ApJ...452..710N} Narayan, R., \& Yi,
   I.\ 1995, \apj, 452, 710 
 \bibitem[Nayakshin et al.(2012)]{2012ApJ...753...15N} Nayakshin, S., Power,
  C., \& King, A.~R.\ 2012, \apj, 753, 15 
 \bibitem[Netzer \& Peterson(1997)]{NaP97}Netzer H., Peterson B.M., 1997, in
  Astronomical Time Series, ed.\ D.\ Maoz, A.\ Sternberg, \& E.M.\ Leibowitz (Dordrecht: Kluwer), 85
 \bibitem[Nguyen et al.(2017)]{2017ApJ...836..237N} Nguyen, D.~D., Seth, A.~C.,
  den Brok, M., et al.\ 2017, \apj, 836, 237 
 \bibitem[Nucita et al.(2017)]{2017ApJ...837...66N} Nucita, A.~A., Manni, L.,
   De Paolis, F., Giordano, M., \& Ingrosso, G.\ 2017, \apj, 837, 66 
 \bibitem[Oka et al.(2016)]{Oka:2016} Oka, T., Mizuno, R., Miura, K.,
   Takekawa, S.\ 2016, ApJ, 816, L7
 \bibitem[Pacucci et al.(2016)]{2016MNRAS.459.1432P} Pacucci, F., Ferrara, A.,
   Grazian, A., et al.\ 2016, \mnras, 459, 1432 
 \bibitem[Pardo et al.(2016)]{2016ApJ...831..203P} Pardo, K., Goulding, A.~D.,
   Greene, J.~E., et al.\ 2016, \apj, 831, 203 
 \bibitem[Pasham et al.(2014)]{2014Natur.513...74P} Pasham, D.~R., Strohmayer,
   T.~E., \& Mushotzky, R.~F.\ 2014, \nat, 513, 74 
 \bibitem[Pasham et al.(2015)]{2015:Pasham} Pasham, D.R., Cenko, S.B., Zoghbi,
  A., Mushotzky, R.F., Miller, J., Tombesi, F.\ 2015, ApJL, 811, L11
 \bibitem[Pastorini et al.(2007)]{2007A&A...469..405P} Pastorini, G., Marconi,
   A., Capetti, A., et al.\ 2007, \aap, 469, 405 
 \bibitem[Paturel et al.(2003)]{Pat03}Paturel G., Petit C., Prugniel P., Theureau G., Rousseau J.,
  Brouty M., Dubois P., \& Cambr{\'e}sy L.\ 2003, A\&A\, 412, 45
 \bibitem[Plotkin et al.(2012)]{plotkin12} Plotkin R.M., Markoff S., Kelly
   B.C., K\"ording E., Anderson S.F. 2012, MNRAS, 419, 267
 \bibitem[Portegies Zwart \& McMillan(2002)]{portegieszwart02} Portegies Zwart S.F., McMillan S.L.W.\ 2002, ApJ, 576, 899
 \bibitem[Pour-Imani et al.(2016)]{2016ApJ...827L...2P} Pour-Imani, H.,
   Kennefick, D., Kennefick, J., et al.\ 2016, \apjl, 827, L2 
 \bibitem[Pringle \& Rees(1972)]{1972A&A....21....1P} Pringle, J.~E., \& Rees,
   M.~J.\ 1972, \aap, 21, 1 
 \bibitem[Press et al.(1992)]{1992nrfa.book.....P} Press, W.~H., Teukolsky,
   S.~A., Vetterling, W.~T., \& Flannery, B.~P.\ 1992, Cambridge: University
   Press, |c1992, 2nd ed.,
 \bibitem[Puerari \& Dottori(1992)]{1992A&AS...93..469P} Puerari, I., \&
   Dottori, H.~A.\ 1992, \aaps, 93, 469 
 \bibitem[Querejeta et al.(2015)]{2015ApJS..219....5Q} Querejeta, M., Meidt,
   S.~E., Schinnerer, E., et al.\ 2015, \apjs, 219, 5 
 \bibitem[Quinlan \& Shapiro(1990)]{1990ApJ...356..483Q} Quinlan, G.~D., \&
  Shapiro, S.~L.\ 1990, \apj, 356, 483 
 \bibitem[Rees(1984)]{1984ARA&A..22..471R} Rees, M.~J.\ 1984, \araa, 22, 471 
 \bibitem[Regan et al.(2017)]{2017NatAs...1E..75R} Regan, J.~A., Visbal, E.,
   Wise, J.~H., et al.\ 2017, Nature Astronomy, 1, 0075 
 \bibitem[Reid et al.(2014)]{2014ApJ...796....2R} Reid, M.~J., McClintock,
  J.~E., Steiner, J.~F., et al.\ 2014, \apj, 796, 2 
 \bibitem[Reines et al.(2013)]{2013ApJ...775..116R} Reines, A.~E., Greene,
   J.~E., \& Geha, M.\ 2013, \apj, 775, 116 
 \bibitem[Reines \& Volonteri(2015)]{2015ApJ...813...82R} Reines, A.~E., \&
   Volonteri, M.\ 2015, \apj, 813, 82 
 \bibitem[Ringermacher \& Mead(2009)]{2009AJ....137.4716R} Ringermacher,
   H.~I., \& Mead, L.~R.\ 2009, \aj, 137, 4716 
 \bibitem[Roediger \& Courteau(2015)]{2015MNRAS.452.3209R} Roediger, J.~C., \&
   Courteau, S.\ 2015, \mnras, 452, 3209 
 \bibitem[Saglia et al.(2016)]{2016ApJ...818...47S} Saglia, R.~P., Opitsch,
   M., Erwin, P., et al.\ 2016, \apj, 818, 47 
 \bibitem[Satyapal et al.(2009)]{satyapal09} Satyapal S., B{\"o}ker T.,
    Mcalpine W., Gliozzi M., Abel N.P., Heckman T.\ 2009, \apj, 704, 439
 \bibitem[Savorgnan(2016)]{2016ApJ...821...88S} Savorgnan, G.~A.~D.\ 2016,
   \apj, 821, 88 
 \bibitem[Savorgnan \& Graham(2015)]{2015MNRAS.446.2330S} Savorgnan, G.~A.~D.,
   \& Graham, A.~W.\ 2015, \mnras, 446, 2330 
 \bibitem[Savorgnan \& Graham(2016)]{2016ApJS..222...10S} Savorgnan, G.~A.~D.,
  \& Graham, A.~W.\ 2016, \apjs, 222, 10 
 \bibitem[Savorgnan et al.(2016)]{2016ApJ...817...21S} Savorgnan, G.~A.~D.,
  Graham, A.~W., Marconi, A., \& Sani, E.\ 2016, \apj, 817, 21 
 \bibitem[Schlafly \& Finkbeiner(2011)]{2011ApJ...737..103S} Schlafly, E.~F.,
  \& Finkbeiner, D.~P.\ 2011, \apj, 737, 103
 \bibitem[Schneider et al.(2002)]{2002ApJ...571...30S} Schneider, R., Ferrara,
  A., Natarajan, P., \& Omukai, K.\ 2002, \apj, 571, 30 
 \bibitem[Schombert(2011)]{2011arXiv1107.1728S} Schombert, J.\ 2011,
   arXiv:1107.1728 
 \bibitem[Schombert et al.(1995)]{1995AJ....110.2067S} Schombert, J.~M.,
   Pildis, R.~A., Eder, J.~A., \& Oemler, A., Jr.\ 1995, \aj, 110, 2067 
 \bibitem[Schwarzschild \& Spitzer(1953)]{1953Obs....73...77S} Schwarzschild,
  M., \& Spitzer, L.\ 1953, The Observatory, 73, 77 
 \bibitem[Scott et al.(2013)]{2013ApJ...768...76S} Scott, N., Graham, A.~W., \&
   Schombert, J.\ 2013, \apj, 768, 76
 \bibitem[Secrest et al.(2012)]{secrest12} Secrest N.J., Satyapal S., Gliozzi
   M., Cheung C.C., Seth A.C., B\"oker T.\ 2012, \apj, 753, 38
 \bibitem[Secrest et al.(2013)]{secrest13} Secrest N.J., Satyapal S., Moran
   S.M., Cheung C.C., Giroletti M., Gliozzi M., Bergmann M.P., Seth A.C. 2013,
   ApJ, 777, 139
 \bibitem[Seigar et al.(2005)]{2005MNRAS.359.1065S} Seigar, M.~S., Block,
   D.~L., Puerari, I., Chorney, N.~E., \& James, P.~A.\ 2005, \mnras, 
 \bibitem[Seigar et al.(2008)]{2008ApJ...678L..93S} Seigar, M.~S., Kennefick,
   D., Kennefick, J., \& Lacy, C.~H.~S.\ 2008, \apjl, 678, L93 
 \bibitem[Seth et al.(2008)]{2008ApJ...678..116S} Seth, A., Ag{\"u}eros, M.,
   Lee, D., \& Basu-Zych, A.\ 2008, \apj, 678, 116-130 
 \bibitem[Shankar et al.(2004)]{2004MNRAS.354.1020S} Shankar, F., Salucci, P.,
  Granato, G.~L., De Zotti, G., \& Danese, L.\ 2004, \mnras, 354, 1020 
 \bibitem[Shannon et al.(2015)]{2015Sci...349.1522S} Shannon, R.~M., Ravi, V.,
  Lentati, L.~T., et al.\ 2015, Science, 349, 1522 
 \bibitem[Shapiro \& Teukolsky(1985)]{1985ApJ...292L..41S} Shapiro, S.~L., \&
   Teukolsky, S.~A.\ 1985, \apjl, 292, L41
 \bibitem[Shih et al.(2003)]{2003MNRAS.341..973S} Shih, D.~C., Iwasawa, K., \&
   Fabian, A.~C.\ 2003, \mnras, 341, 973 
 \bibitem[Sijacki et al.(2007)]{2007MNRAS.380..877S} Sijacki, D., Springel,
   V., Di Matteo, T., \& Hernquist, L.\ 2007, \mnras, 380, 877
 \bibitem[Soria et al.(2010)]{Soria:2010}Soria, R., Hau, G.K.T., Graham, A.W.,
   Kong, A.K.H., Kuin, N.P.M., Li, I.-H., Liu, J.-F. \& Wu, K.\ 20\ 10, MNRAS,
   405, 870
 \bibitem[Soria et al.(2017)]{soria17} Soria R., Musaeva A., Wu K., Zampieri
   L., Federle S., Urquhart R., van der Helm E., Farrell S.A.\ 2017, MNRAS,
   469, 886
 \bibitem[Spera et al.(2015)]{spera15} Spera M., Mapelli M., Bressan A., 2015,
   MNRAS, 451, 4086
 \bibitem[Spera \& Mapelli(2017)]{spera17} Spera M., Mapelli M., 2017, MNRAS,
   470, 4739
 \bibitem[Stone \& Metzger(2016)]{2016MNRAS.455..859S} Stone, N.~C., \&
   Metzger, B.~D.\ 2016, \mnras, 455, 859 
 \bibitem[Strader et al.(2012)]{strader12} Strader J., Chomiuk L., Maccarone
   T.J., Miller-Jones J.C.A., Seth A.C., Heinke C.O., Sivakoff G.R.\ 2012,
   ApJ, 750, L27
 \bibitem[Sutton et al.(2012)]{2012MNRAS.423.1154S} Sutton, A.~D., Roberts,
   T.~P., Walton, D.~J., Gladstone, J.~C., \& Scott, A.~E.\ 2012, \mnras, 423,
   1154
 \bibitem[Swartz et al.(2008)]{swartz08} Swartz D.A., Soria R., Tennant A.F.,
   2008, ApJ, 684, 282
 \bibitem[Tadhunter et al.(2003)]{2003MNRAS.342..861T} Tadhunter, C., Marconi,
   A., Axon, D., et al.\ 2003, \mnras, 342, 861
 \bibitem[T{\'a}pai(2015)]{2015mgm..conf..957T} T{\'a}pai, M.~K., Zolt{\'a}n
   Gergely, L{\'a}szl{\'o} {\'A}.\ 2015, Thirteenth Marcel Grossmann Meeting,
   eds. Rosquist Kjell et al., Published by World Scientific Publishing, 957
   (arXiv:1212.4973)
 \bibitem[Taylor et al.(2011)]{2011MNRAS.418.1587T} Taylor, E.~N., Hopkins,
   A.~M., Baldry, I.~K., et al.\ 2011, \mnras, 418, 1587 
 \bibitem[Terashima et al.(2015)]{terashima15} Terashima Y., Hirata Y., Awaki H., Oyabu S., Gandhi P., Toba Y., Matsuhara H.\ 2015, \apj, 814, 11 
 \bibitem[Tremaine et al.(2002)]{2002ApJ...574..740T} Tremaine, S., Gebhardt,
   K., Bender, R., et al.\ 2002, \apj, 574, 740 
 \bibitem[Turner(1991)]{1991AJ....101....5T} Turner, E.~L.\ 1991, \aj, 101, 5
 \bibitem[Umeda \& Nomoto(2003)]{2003Natur.422..871U} Umeda, H., \& Nomoto,
  K.\ 2003, \nat, 422, 871 
 \bibitem[Umemura et al.(1993)]{1993ApJ...419..459U} Umemura, M., Loeb, A., \&
   Turner, E.~L.\ 1993, \apj, 419, 459 
 \bibitem[Watson et al.(2009)]{watson09} Watson M.G., et al.\ 2009, A\&A, 493, 339
 \bibitem[Webb et al.(2010)]{webb10} Webb N.A., Barret D., Godet O., Servillat
   M., Farrell S.A., Oates S.R.\ 2010, \apjl, 712, L107
 \bibitem[Webb et al.(2014)]{webb14} Webb N.A., Godet O., Wiersema K., Lasota
   J.-P., Barret D., Farrell S.A., Maccarone T.J., Servillat M.\ 2014, \apjl,
   780, 9
 \bibitem[Webb et al.(2017)]{webb17} Webb N.A., et al.\ 2017, \aap, 602, A103
 \bibitem[Willmer(2018)]{Will18}Willmer, C.N.A.\ 2018, ApJS, submitted
   (arXiv:1804.07788)
 \bibitem[Yan et al.(2015)]{2015ApJ...811...23Y} Yan, Z., Zhang, W., Soria,
   R., Altamirano, D., \& Yu, W.\ 2015, \apj, 811, 23
 \bibitem[Yu et al.(2018)]{2018arXiv180606591Y} Yu, S.-Y., Ho, L.~C., Barth,
   A.~J., \& Li, Z.-Y.\ 2018, arXiv:1806.06591 
 \bibitem[Zel’dovich \& Podurets(1965)]{ZaP1965}Zel’dovich, Ya. B., and
  Podurets, M. A. 1965, Astr. Zh., 42, 963 (English translation in Soviet Astr.—AJ, 9, 742) (ZP).
 \bibitem[Zhong et al.(2015)]{2015ApJ...811...22Z} Zhong, S., Berczik, P., \&
   Spurzem, R.\ 2015, \apj, 811, 22 
 \bibitem[Zolotukhin et al.(2016)]{2016ApJ...817...88Z} Zolotukhin, I., Webb,
    N.~A., Godet, O., Bachetti, M., \& Barret, D.\ 2016, \apj, 817, 88 

\end{thebibliography}
\end{document}